\documentclass[aps,prx,twocolumn,showpacs,superscriptaddress,longbibliography]{revtex4-2}
\usepackage[english]{babel}
\usepackage[utf8]{inputenc}
\usepackage{graphicx}
\usepackage[unicode=true,pdfusetitle,
 bookmarks=true,bookmarksnumbered=false,bookmarksopen=false,
 breaklinks=false,pdfborder={0 0 0},pdfborderstyle={},backref=false,colorlinks=true]
 {hyperref}
 \hypersetup{
 citecolor=blue,
 urlcolor=blue,
 linkcolor=blue}

 \usepackage{latexsym,amsmath,verbatim}
\usepackage{amssymb} 
\usepackage{amsfonts}
\usepackage{colortbl}
\usepackage{array}
\usepackage{stackrel}
\usepackage{bm}
\usepackage{nicefrac}
\usepackage{rotating}
\usepackage{float}
\usepackage[final]{pdfpages}
\usepackage{lipsum}  
\usepackage{braket}
\usepackage{dsfont}
\usepackage[normalem]{ulem}
\usepackage{xcolor}
\newcommand{\pv}[1]{{{#1}}}

\newcommand{\new}[1]{{\color{black} {#1}}}

\makeatletter
\AtBeginDocument{\let\LS@rot\@undefined}
\makeatother

\begin{document}

\title{Multi-scale Laplacian community detection in heterogeneous networks}
%in the framework of the Laplacian Renormalization Group}

%\title{Laplacian Renormalization Group: Community detection and Markov stability}

\author{Pablo Villegas}
\affiliation{`Enrico Fermi' Research Center (CREF), Via Panisperna 89A, 00184 - Rome, Italy}
\author{Andrea Gabrielli}
\email[Corresponding author: andrea.gabrielli@uniroma3.it]{}
\affiliation{`Enrico Fermi' Research Center (CREF), Via Panisperna 89A, 00184 - Rome, Italy}
\affiliation{Dipartimento di Ingegneria Civile, Informatica e delle Tecnologie Aeronautiche, Università degli Studi `Roma Tre', Via Vito Volterra 62, 00146 - Rome, Italy}
\author{Anna Poggialini}
\affiliation{`Enrico Fermi' Research Center (CREF), Via Panisperna 89A, 00184 - Rome, Italy}
\affiliation{Dipartimento di Fisica Universit\`a ``Sapienza”, P.le
  A. Moro, 2, I-00185 Rome, Italy.}

\author{Tommaso Gili}
\affiliation{Networks Unit, IMT Scuola Alti Studi Lucca, Piazza San Francesco 15, 55100- Lucca, Italy.}
\affiliation{Institute for Complex Systems (ISC), CNR, UoS Sapienza, Piazzale Aldo Moro 2, 00185 - Rome, Italy.}

\vspace{0.5cm}

\vspace{0.5cm}

\renewcommand\refname{} % required to avoid the bibliography heading

\begin{abstract}

%The intertwined nature of complex systems, encoded in rich network structures, represents a large class of real-world interacting agents. A fundamental problem involves finding inherent partitions, clusters, or communities. We scrutinize inter-node communicability by taking advantage of information diffusion pathways throughout complex networks based on recent renormalization group approaches to shed further light on these issues. Our definition provides a unifying framework for multiple partitioning measures: we show that Markov stability and Laplacian spectrum methods are two sides of the same coin.

\new{Heterogeneous and complex networks represent intertwined interactions between real-world elements or agents. Determining the multi-scale mesoscopic organization of clusters and intertwined structures is still a fundamental and open problem of complex network theory. By taking advantage of the recent Laplacian Renormalization Group, we scrutinize information diffusion pathways throughout networks to shed further light on this issue. Based on inter-node communicability, our definition provides a clear-cut framework for resolving the multi-scale mesh of structures in complex networks, disentangling their intrinsic arboreal architecture. As it does not consider any topological null-model assumption, the LRG naturally permits the introduction of scale-dependent optimal partitions. Moreover, we demonstrate the existence of a particular class of nodes, called {\em metastable} nodes, that switching regions to which they belong at different scales, likely playing a pivotal role in cross-regional communication and, therefore, in managing macroscopic effects of the whole network.}

\end{abstract}

\maketitle
%%%%%%%%%%%%%%%%%%%%%%%%%%%%%%%%%%%%%%%%%%%%%%%%%%%%%%%%%%%%%%%%

Community structures emerge as a ubiquitous feature of real-world networks. Detection of relevant network substructures \pv{is paramount} to correctly understand their often hidden multiple mesoscopic scales and particular functionalities \cite{wagner2007,Newman}. The level of compartmentalization of a system, \emph{the modularity}, has been proposed as a generic requirement, or an optimal solution, for a system in a changing environment to be evolvable \cite{Lorenz}. High modularity, therefore, pervades biology on different scales from proteins and genes \cite{han2004} to cells \cite{norman2013,gavin2006}, the human interactome \cite{taylor2009} or ecosystems \cite{Villegas2021}. Modules form the basis of transcriptional regulatory networks \cite{babu2004}, have been found at every scale in metabolic networks \cite{ravasz2002}, and have been hypothesized to confer a high efficiency of information transfer between nodes at low connection cost in brain networks \cite{bullmore2012}.

%Classification of Modularity algorithms
In a seminal paper, Girvan and Newmann \cite{Newman} proposed a celebrated algorithm by progressively removing edges from the original network. The inherent nature of the algorithm led them to introduce a quantitatively ``stopping criterion": the \emph{modularity, Q}, which quickly became the fundamental ingredient of many clustering methods \cite{Newman2006}. \pv{Reichardt and Bornholdt simultaneously proposed an analogous measure as a solution of a q-state Potts model, demonstrating its formal equivalence to Girvan and Newman's algorithm \cite{Reich2004, Reich2006}.} Modularity maximization is one of the most widely used algorithms to detect communities in a network. However, an exhaustive modularity maximization implies an NP-complete problem, whose complexity increases exponentially with the total number of nodes, making it practically unfeasible for most cases of interest (we refer to \cite{Fortunato2010, Fortunato2016} for an extensive discussion on the issue).

More sophisticated procedures rely on the spectral properties of the graph Laplacian, which encode the network diffusion modes (i.e., allowing to explore the network momentum space \cite{LRG}). Indeed, this is the basis for one of the best-known \pv{graph partitioning methods}: spectral partitioning \cite{Fiedler1973,Pothen1990}. A related elegant proposal combines spectral methods with clustering techniques, projecting the network nodes into an eigenvector space of variable (tunable) dimensionality \cite{Donetti2004}. Note that all the algorithms taking~ advantage of spectral methods, which operate in Fourier space, need an $N$-dimensional space due to a lack of well-defined spatial embedding of the network. Another set of methods is based on the exploration properties of random walks on top of the network.  In a general random walk on a graph, the probability of leaving a vertex is distributed among the outgoing edges according to their weight \cite{Barahona2010}. This implies that the intrinsic network topological heterogeneities can make a random walker to get trapped in a subnetwork for a relatively long time before leaving it (see, e.g., \cite{Masuda2017} for an excellent review on the field). This provides an alternative definition of community. Hence, the concept of \emph{Markov stability} has been introduced \cite{Barahona2010}, which quantifies the tendency for a random walker to stay inside a community for a long time \cite{Masuda2017, Patelli2020}, opening the door to additional quality measures of a graph partition based on dynamical concepts, therefore ranking partitions and establishing their relevance over time scales.

Detecting mesoscale structures (i.e., communities) in networks has become a fundamental problem --and still a matter of debate \cite{Fortunato2022}-- in network science, where many different methods have been proposed and classified according to their performance \cite{LFComm}. However, the problem lacks a holistic perspective, \new{i.e., a unified theoretical framework} capable of setting the limitations of the different methods. \pv{For instance, modularity has been demonstrated to be a linear approximation of Markov stability \cite{Lambiotte2014}. Besides, community detection algorithms based on quality functions strongly depend on the null models they must consider to meaningfully quantify their statistical significance \cite{Lambiotte2014, Expert2011}. Another crucial problem is that} the definition of a community in general is strictly connected \pv{to functional properties and, thus,} to the process on the network for which we want to find the optimal partition \cite{Rosvall2014}. Moreover, the proper identification of local communities represents a fundamental open problem when tackled through usual community detection methods \cite{Fortunato2022}. In a nutshell, identifying basic structural and dynamic \pv{(functional)} communities \new{disentangling the arboreal architecture of a complex network} \pv{without any topological null-model assumption,} raises a general problem of definition that still needs to be resolved.

%In addition, it has been recently shown how considering different dynamical processes in various integrated systems may lead to significant consequences for community detection, ranking, and information spreading 

In all cases, tackling the root of the problem requires \pv{considering} the evolution of inter-node communication throughout multiple scales. In classical statistical physics, the Renormalization Group (RG) --one of the most important concepts and tools in theoretical physics of the second half of the 20th century \cite{Fisher1974,Wilson1974,Amit,Kardar}-- allows for connecting extremely varied spatiotemporal scales of translationally invariant physical systems embedded in a homogeneous (typically Euclidean) space. Recently, a natural extension of this concept to heterogeneous networked spaces, the Laplacian Renormalization Group (LRG), grounding on inherent information dynamics, has been introduced. It consequently appears as the natural framework to define and detect complex network mesoscales and structures \cite{LRG}. As the theory of LRG acts as a 'zoom lens' for heterogeneous structures, \pv{tightly coupling temporal and spatial scales by means of the network Laplacian,} it allows us \pv{to scrutinize} all the possible diffusion paths at different --and arbitrary-- network resolution scales.

\new{Here, grounding on the LRG, we unravel the entire multi-scale structure of every network at different resolution levels. In particular, we reconstruct the hierarchical network dendrogram (the arboreal architecture of the network), proposing a new concept of communication distances among nodes by taking advantage of the LRG in their $k-space$ formulation without needing \emph{a priori} null models or statistical inference. Therefore, we show how the optimal partitions are those determined by the slowing down of the information diffusion throughout the network. In particular, we evidence the existence of metastable nodes that can dynamically switch the region to which they belong at different intrinsic network timescales. Finally, we illustrate the emergence of functional dendrograms when a set of random walks explore the network structure and discuss the bridges with previous frameworks, such as modularity maximization, Markov stability, and spectral methods, opening new avenues in future developments in the networks field.}

\section{Real-Space distances of non-euclidean topological structures}

\new{We ground our approach on the recently proposed LRG \cite{InfoCore, LRG}, and in the Laplacian density matrix \cite{Domenico2016, Ghavasieh2020},
\begin{equation}
 \hat \rho(\tau)=\frac{e^{-\tau \hat L}}{Tr (e^{-\tau \hat L})}\,,
 \label{RhoMat}
\end{equation}

 where $\hat L=\hat D- \hat A$ represents the symmetric Laplacian, $\hat{D}$ is the degree matrix and $\hat A$ the adjacency matrix of the network. In particular, the entropic susceptibility (or specific heat, see Appendix \ref{StatPhys}) $C(\tau)=-\frac{dS}{d \log \tau}$, allows us to analyze the multi-scale organization of any network accurately \cite{InfoCore, LRG}. The operator $\hat K(\tau)=e^{-\tau\hat L}$ or, equivalently $\hat \rho (\tau)$, gives a robust measure of \emph{information communicability} or diffusion strength between pair of nodes on a time scale $\tau$, accounting for the sum of diffusion trajectories along all possible paths connecting nodes $i$ and $j$ in a temporal scale $\tau$ \cite{Masuda2017, Moretti2019}. }
\begin{figure}[hbtp]
    \centering
    \includegraphics[width=1.0\columnwidth]{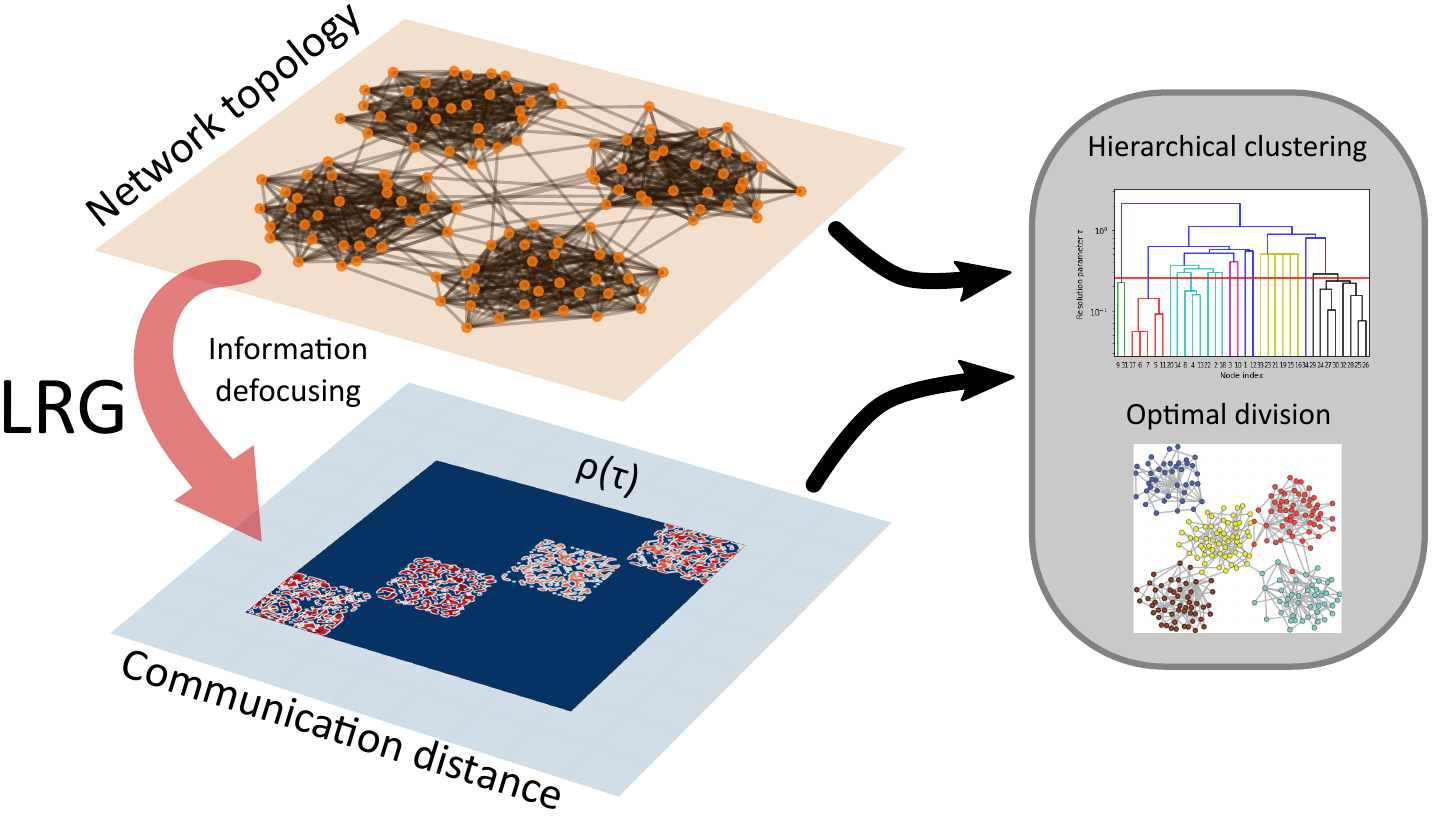}
    \caption{\textbf{Defining multi-scale Kadanoff supernodes.} \emph{Communication
distance} between nodes reflects the underlying hidden complex topology, capturing the inter-nodes communicability and giving a natural merging of nodes at specific times, $\tau$. The LRG gives a natural interpretation of the network scales in terms of network eigenmodes, thereby enabling a deeper understanding of the mesoscopic properties of the network.}
    \label{fig:my_label}
\end{figure}

%%%%%%%%%%%%%%%%%%%%%%%
%We point out that the peculiar structural organization of the network relies, however, on the Laplacian matrix, which encodes the intrinsic topological properties and determine coarse graining and renormalization group procedures on top of heterogeneous structures \cite{LRG}. In particular, the Laplacian governs the diffusion dynamics on the network through the Laplace$/$heat equation ($\textbf{s}(\tau)$ is for instance the information distribution on the network nodes)
%\[ 
%\dot{\textbf{s}}(\tau)= - \hat L \textbf{s}(\tau)
%\]
%whose general solution is $\textbf{s}(\tau)=e^{-\tau\hat L}\textbf{s}(0)$. This permits to introduce the concept of Laplacian density matrix as \cite{Domenico2016, Ghavasieh2020},
%\begin{equation}
 %\hat \rho(\tau)=\frac{e^{-\tau \hat L}}{Tr (e^{-\tau \hat L})},
 %\label{RhoMat}
%\end{equation}
%which opened the door to a thorough analysis of network structures through the introduction of the Laplacian entropy \cite{Domenico2016}, $S(\tau)=-\frac{1}{\log N} \mbox{Tr}[\hat \rho (\tau) \log \hat \rho (\tau)]$, and entropic susceptibility (or specific heat) \cite{InfoCore, LRG}, $C(\tau)=-\frac{dS}{d \log \tau}$. In particular, $C(\tau)$ can be seen as a microscope with resolution parameter $\tau$  able to reveal the network characteristic scales. The same quantity also makes  possible to scrutinize the scale-invariant properties of a network \cite{LRG}, linking its amplitude with the effective spectral dimension of the network (see also Appendix \ref{StatPhys}).

\new{First, we propose a \emph{Communication distance} at time $\tau$. From the physical meaning of $e^{-\tau \hat L}$ it is natural to define such symmetric distance between any pair of nodes $i$ and $j$ as $\mathcal{D}_{ij}(\tau)=\frac{1-\delta_{ij}}{K_{ij}(\tau)}$ (or equivalently $\mathcal{D}_{ij}(\tau)=\frac{1-\delta_{ij}}{\rho_{ij}(\tau)}$). 
Indeed, $K_{ij}(\tau)$ gives the fraction of initial information on node $i$ that is transferred to node $j$ by diffusion in a time $\tau$ and {\em vice-versa}. Equivalently, one can write $K_{ij}(\tau)=\frac{\delta s_j(\tau)}{\delta s_i(0)}$, where $\delta s_j(\tau)$ is the fluctuation of the information on node $j$ at time $t$ induced by a fluctuation of the initial information $\delta s_i(0)$ on node $i$, keeping as fixed the initial condition on the other nodes. Hence, the larger $\mathcal{D}_{ij}(\tau)$, the less the communicability on a time $\tau$ between nodes $i$ and $j$. %, given a fluctuation $\delta s_j(0)$ of the initial information on vertex $j$ keeping fixed the rest of the initial distribution, returns the ration between the fluctuation $\delta s_i(\tau)$ induced on the information on vertex $i$ after a time $\tau$. Therefore, the smaller $-\log[\rho_{ij}(\tau)$ the more communicating the nodes on this time scale. 
It has to be emphasized that $\mathcal{D}_{ij}(\tau)$ defines an ultrametric distance, as it satisfies the following conditions at all $\tau>0$: (i) $\mathcal{D}_{i j} \geq 0$ with $\mathcal{D}_{i j}=0$ {\em if} $i=j$; (ii) $\mathcal{D}_{i j}=\mathcal{D}_{ji}$, and (iii) $\mathcal{D}_{i j} \leq \max \left(\mathcal{D}_{i z}, \mathcal{D}_{z j}\right)$ $\forall z\in {\cal V}$ (where ${\cal V}$ is the set of nodes of the network).}

Even though complex networks lack an explicit spatial embedding, $\mathcal{\hat D}$ allows us to reduce our problem to traditional clustering problems in a metric space \cite{Kaufman90} allowing for the construction of a dendrogram at arbitrary $\tau$ \footnote{In fact, $\mathcal{\hat D}$ is a projection of how points (nodes) are distributed in the N-dimensional k-space representation of the network.}.
To define a stable algorithm for the construction of the dendrogram, we choose the average group clustering algorithm, a compromise between the sensitivity to outliers of the complete clustering and the tendency of single-linkage clustering to inhibit compact clusters \cite{Shao2007, Kaufman90}. In this way, the merging of two clusters happens when the mean distance $\overline{\mathcal{D}}$ between any node belonging to the first cluster and any node belonging to the second cluster becomes smaller than the chosen threshold $\Delta$. The output of this algorithm will thus be a hierarchical tree for each resolution scale $\tau$, without null-model assumptions.

At this point, three natural questions arise: (1) Given the dendrogram at diffusion time $\tau$, is there an optimal partition or cut? (2) Since the metric $\mathcal{\hat D}(\tau)$ may, at different $\tau$, determine different dendrograms with variable composition and length of branches (such as in Markov stability \cite{Barahona2010}), is there an optimal choice of $\tau$? (3) Is it stable the optimal cut at varying $\tau$? We face now the first two questions, leaving the third one in a section below.
%Note however that at different $\tau$ the metric $\mathcal{\hat D}\left[\hat\rho(\tau)\right]$ can, in general determine different dendrograms in which the specific length of branches and their composition can change. Consequently, $\mathcal{\hat D}\left[\hat\rho(\tau)\right]$ induces diverse structural partitions of the network at different times (this situation is shared by other metrics; e.g. the same applies to Markov stability \cite{Barahona2010}). A natural question arises: how to reconcile the identification of a single optimal partition with preserving the multiple intrinsic timescales stemming from the topological heterogeneity of the network?

%Finally, for the sake of clarity, we consider the relative length of the dendrogram by the standardization of the distances $\mathcal{D}'=\mathcal{D}/\mathcal{D}_{\max}$ where $\mathcal{D}_{\max}$, within our choice of the average clustering, is at a given $\tau$ the maximal mean distance between nodes belonging to different clusters. This results in a standardized dendrogram reflecting the intrinsic network mesoscopic structures with a minimum resolution fixed by the scale $\tau$ \footnote{Note that the maximum communication distance can change but we are interested in the relative ordering of the network at different resolution scales}.

\begin{figure}[hbtp]
    \centering
    \includegraphics[width=1.0\columnwidth]{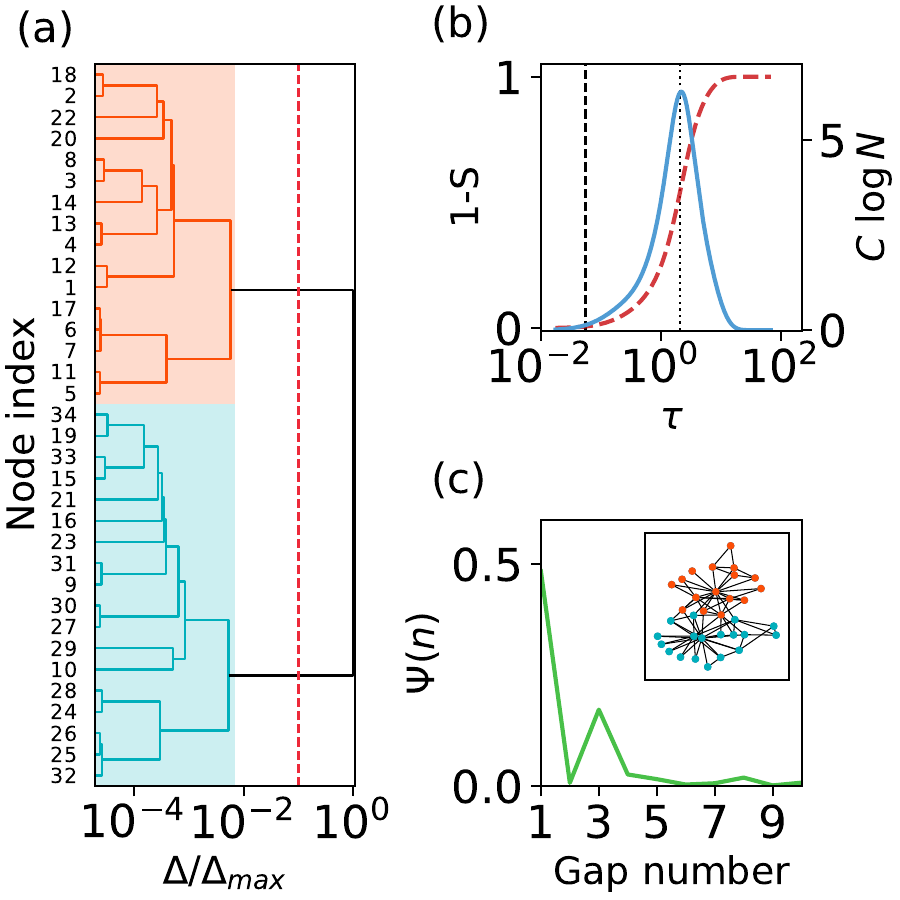}
    \caption{\textbf{LRG multi-scale decomposition.} (a) Normalized dendrogram for the Zachary's karate club with $\tau\lesssim\tau^\prime=1/\lambda_{max}$. The Red dashed line reflects the optimal division as stated by the maximum of $\Psi$. Different communities are shaded in different colors. (b) Entropy parameter (red dashed line, $(1 - S)$), and specific heat (blue solid line, $C$), versus the temporal resolution parameter of the network, $\tau$. Black dashed line indicates $\tau'=1/\lambda_{\max}$. The black dotted line indicates the time of the maximum heat capacity, $(\tau^*)$. (c) Partition Stability Index $(\Psi)$ versus gap number for $\tau\lesssim\tau^\prime$. Note the high $\Psi$ values giving place to the usual division in two and four communities of Zachary's karate club. Inset: Optimal division into communities of the network.}
    \label{Zach}
\end{figure}

Figure \ref{Zach}(a) shows the hierarchical dendrogram for Zachary's karate club \cite{Newman, Zachary}, which serves as a benchmark showing the usual partition of the network in two communities. Figure \ref{Zach}(b) shows the entropic phase transition \cite{InfoCore} for Zachary's karate club  using the Laplacian entropy $S(\tau)$ in time (red dashed line) and of the entropic susceptibility (i.e., the specific heat) $C(\tau)$.

We propose a comparative standard for selecting the optimal cut quantifying the \emph{quality} of the different network partitions encrypted in every dendrogram at a fixed diffusion time $\tau$. In particular, we introduce the Partition Stability Index $\Psi(n;\tau)$, as the ratio between the dendrogram gaps and the total dendrogram length, logarithmically measured. Then,
%(i.e., cuts at different levels of $\Delta$)
\begin{equation}
\Psi(n;\tau)=\mathcal{N}[\log\Delta_{n}(\tau)-\log\Delta_{n+1}(\tau)]\,,
\label{psi-n}
\end{equation}
%$\Psi(n)=\frac{\log\mathcal{D}_{n+1}-\log\mathcal{D}_n}{\log\mathcal{D}_{1}-\log\mathcal{D}_{N}}$,
where $\Delta_n(\tau)$ is the value of the threshold $\Delta$ corresponding to the $n^{th}$ dendrogram branching, starting from the network as a single cluster (with this choice of the index $n$, $\Delta_n(\tau)$ is a decreasing function of $n$) and $\mathcal{N}=[\log\Delta_{1}(\tau)-\log\Delta_{n_{max}}(\tau)]^{-1}$ is a normalization constant.

\new{The quantity $\Psi(n;\tau)$ measures the minimal distance (dissimilarity) in logarithmic scale between all the pairs of clusters belonging to level $n+1$ and to level $n$. The logarithmic scale is motivated by the definition of the entropy $(S(\tau) \sim \left<\log \hat \rho(\tau)\right>_{\tau})$ itself: the higher $\Psi(n;\tau)$, the higher the (Laplacian) information gap between two consecutive branching indices $n$ and $n+1$. It is, therefore, natural to consider the optimal cut of the dendrogram at the value $n$ maximizing $\Psi(n;\tau)$, which contains the maximal Laplacian information induced by the network's topology. In the case of multiple intrinsic scales of the network, $\Psi(n;\tau)$  can present different local maxima, reflecting characteristic mesoscopic network regions.}

\new{The interpretation of the dendrogram at different $\tau$'s needs of the recently introduced theory of the LRG \cite{LRG}. It grounds on the fact that $\hat L$ --and specifically its eigenvalues $\lambda$ and eigenvectors $\ket{\lambda}$-- encodes the whole information about all the independent network diffusion modes (see Appendix \ref{LRGAp} and Appendix \ref{CDetection}).} As shown in \cite{InfoCore} and \cite{LRG}, these modes collected in the operator $e^{-\tau\hat L}$ determine the clusters of nodes that are informationally connected at the time-scale $\tau$. Indeed, the LRG takes advantage of the fact that, varying the diffusion time $\tau$, the operator $\hat\rho(\tau)$ acts as a sort of ``scanner" of the characteristic network scales \cite{klemm2023}. Such mesoscopic scales can be detected by progressively integrating out smaller and smaller Laplacian eigenmodes and, consequently, the corresponding network structures as $\tau$ increases. For finite but large networks, the susceptibility $C(\tau)$, determines the existence of a resolution window where the different partitions of the network in clusters are relevant. The smallest possible network scale corresponds to the maximum eigenvalue $\lambda_{max}$: the choice of a diffusion time-scale $\tau\sim \nicefrac{1}{\lambda_{max}}$ coincides with the finest possible resolution of the network structures (this sheds light on the previously enigmatic origin of the resolution limit on the community detection field \cite{Fortunato2007}). Starting from this scale $(\tau'=1/\lambda_{max})$, one can consider progressively larger diffusion times integrating out smaller network eigenmodes up to $\tau\sim 1/\lambda_{gap}$, where $\lambda_{gap}$ is the smallest positive eigenvalue, also called Fiedler eigenvalue or {\em gap}, and monitoring the stability across diffusion scales of the partition of the network into clusters (i.e., defocusing microscopic details). This is related to the maximal characteristic network scale marked by the specific heat peak at long times $(\tau^*)$, beyond which it rapidly decreases to zero.

%, allowing us to propose a stringent criterion to choose the dendrogram for a particular time $\tau'$

The usual definition of \emph{modularity} $Q$ given in Eq.\eqref{ModulAp} (see \cite{Blondel2008, Newman2004}), can be equivalently rewritten as $Q=\stackrel[i]{}{\sum}\left(e_{ii}-a_i^2\right)$ \cite{Clauset2004} providing a `quality' function of the network partition in communities indicated by the index $i$, where $e_{ij}$ is the fraction of edges with one end vertices in community $i$ and the other in community $j$ and $a_i=\sum_j e_{ij}$.  It measures the difference between the empiric fraction of edges connecting nodes inside the same community and the expected one by randomizing and conserving the degree sequence. This implies that it does not take into account other possibly important topological features of the network (e.g., loops or cycles \cite{Dorogovstev2008} or the time scale separation in linear network dynamics \cite{lambiotte_schaub_2022}) that may have a profound impact on the optimal partition of nodes in functional units when for instance a dynamical system is defined on top of the network. \pv{We emphasize that modularity maximization always assumes a statistical inference perspective to detect any modular architecture (in particular, using the configuration model or the Erd\H{o}s-R\'enyi as null models, as shown in \cite{Lambiotte2014}).}

Figure \ref{Zach}(c) shows the behavior of $\Psi(n)$ for Zachary's karate club and the inset the optimal partition for the time $\tau$ indicated with the vertical dashed line in  Fig.~\ref{Zach}(b). We have also analyzed the structural organization of two paradigmatic hierarchical nested networks: hierarchical-modular networks with a preferential attachment rule (HM-CP, producing a core-periphery structure involving central connector hubs with local and global rich clubs \cite{Zamora2016}) and the Dorogovtsev-Goltsev-Mendes graph (DGM), a \emph{pseudo-fractal} graph with high clustering properties \cite{Dorogovstev2002}. As better illustrated in Appendix \ref{HMCP}, our LRG-based approach can capture the full intrinsic modular structure of nested hierarchical networks (HM-CP), even when the inter-module connections and communication paths tend to be centralized through the hubs, as observed in real neural and brain networks \cite{Zamora2019}. The same happens for the highly hierarchical DGM graph, whereas usual community detection methods present severe issues in unraveling static communities being unable to disentagle the complete DGM hierarchical structure (see Appendix \ref{HMCP}). Figures \ref{NiceM}(a) and (b) illustrate the emergent nested structure hinted in the distance matrix, $\mathcal{\hat D}$, for HM-CP and DGM networks. Finally, we have tested the Lancichinetti-Fortunato-Radicchi benchmark \cite{LFR}, a network generator with {\em a priori} known heterogeneous communities, widely used as a stringent criterion to compare different community detection algorithms (see Appendix \ref{HMCP}).

\begin{figure}[hbtp]
    \centering
    \includegraphics[width=0.8\columnwidth]{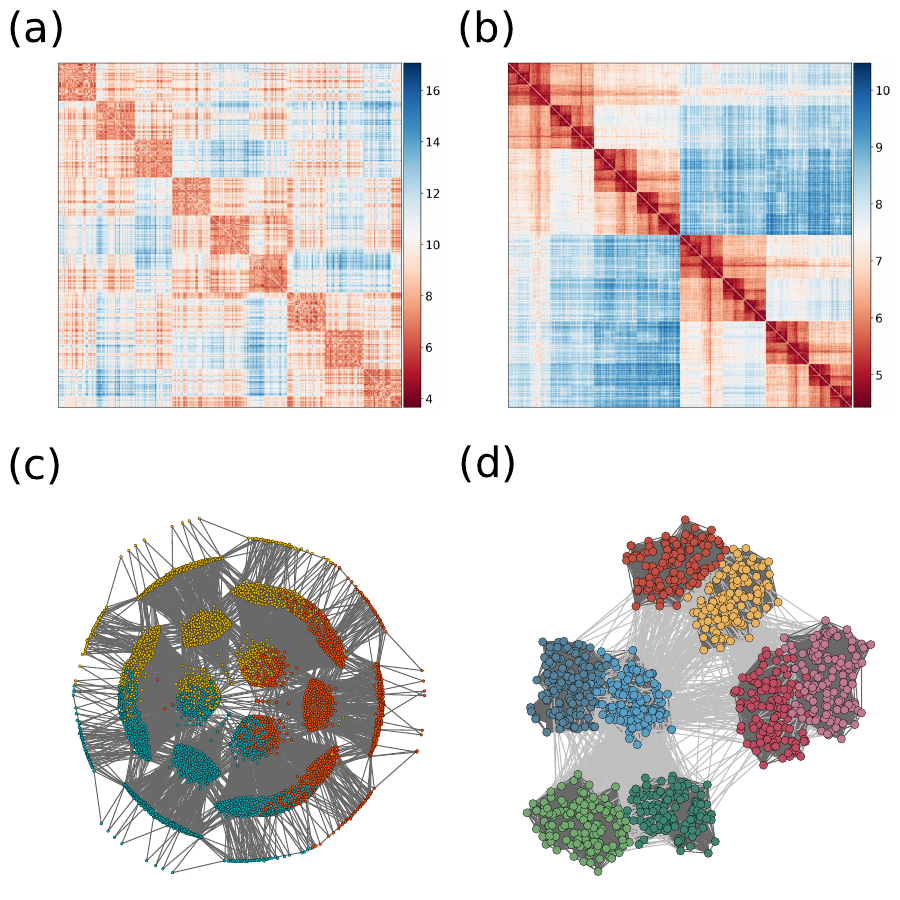}
    \caption{\textbf{Nested structures.} Logarithm of the communication distance at $\tau'=1/\lambda_{max}$ for: \textbf{(a)} a DGM graph, and \textbf{(b)} a hierarchic modular network with core-periphery structure (HM-CP). \textbf{(c)} Division into 3 communities for the DGM case and \textbf{(d)} division into 8 communities for a HM-CP network. }
    \label{NiceM}
\end{figure}

\section{Local modularity in heterogeneous networks}

A recent paper \cite{Fortunato2022} raised central questions about how to detect clusters of nodes of a network playing an important role on the local scale, but that cannot be detected by global optimization methods like modularity maximization.
\pv{Indeed, in many real networks, functionally important clusters are only linked to a small fraction of other clusters on a local scale, so global criteria can completely overshadow them}. For instance, the coexistence of this sort of local modularity with global nestedness is a common key ingredient to ensure evolutionary stability in host-pathogen infection networks \cite{Valverde2020}, while the temporal lobe of the human brain is organized into spatially compact functional modules at the micro-scale \cite{chapeton2022}. Quantifying the local stability across different network scales of these structures is particularly challenging when we only consider an optimization function (e.g., modularity to find the network's internal structure).
This justified the introduction of the concept of {\em localized modularity} \cite{Muff2005}.  

\begin{figure}[hbtp]
    \centering
    \includegraphics[width=1.0\columnwidth]{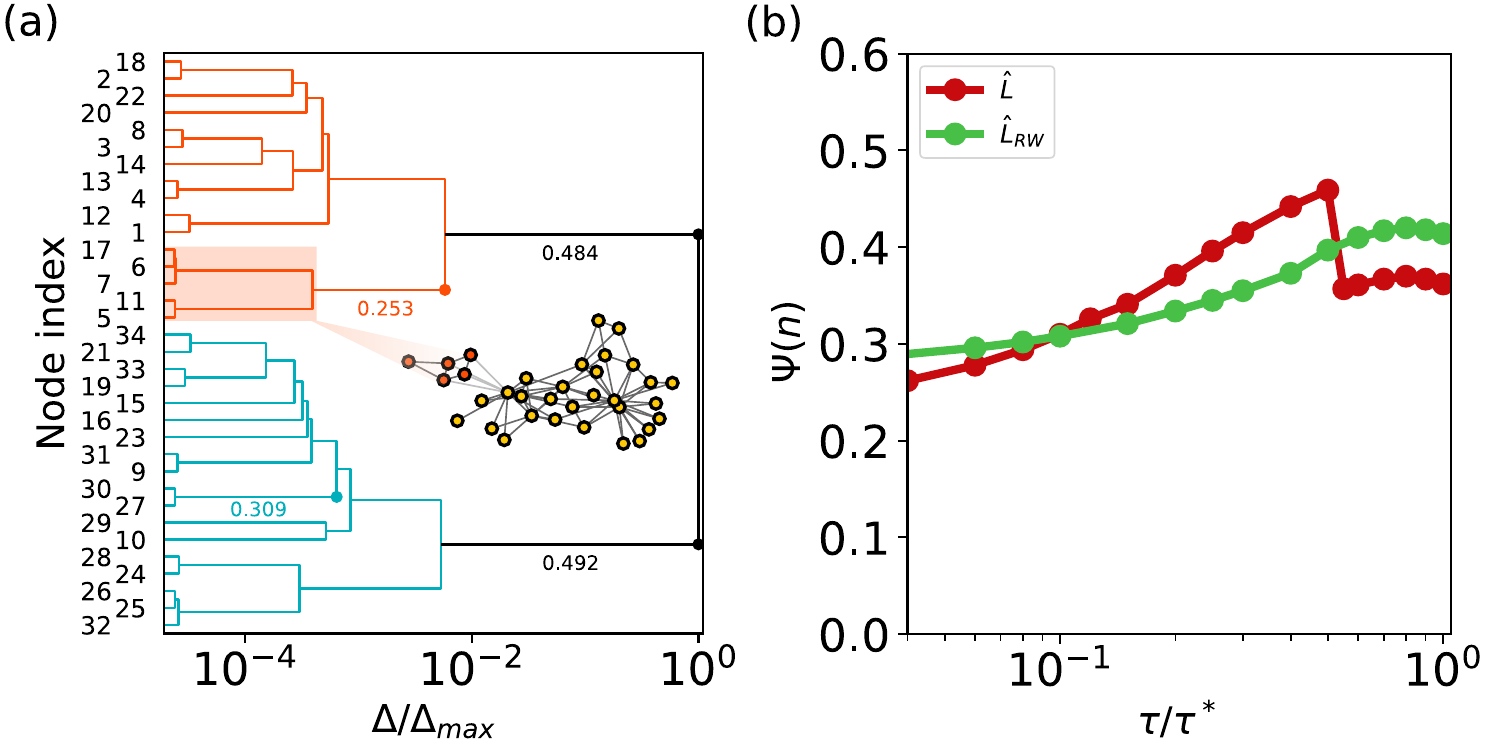}
    \caption{\textbf{Local modularity. (a)} Local modularity in the normalized dendrogram for Zachary's karate club with $\tau\lesssim\tau^\prime=1/\lambda_{max}$. Different relevant values of $(\Psi_\ell)$ have been highlighted. \textbf{(b)} Local Partition Stability Index versus $\tau/\tau^*$ (with $\tau^*$ corresponding to the maximum in C) for the local community highlighted in orange color for both the Laplacian (red line) and the Laplacian RW (green line, see below).}
    \label{Local}
\end{figure}

Since our new theoretical framework depends on a free parameter $\tau$ directly related to the scale of internal structures, it provides a natural solution for a systematic approach to this issue. 
We can, thus, introduce the Local Partition Stability Index $\Psi_L(\tau)$ for each local cluster in the dendrogram at time $\tau$.
It is defined as the length in terms of $\log \Delta$ of the branch of the dendrogram between the creation of a cluster by the merging of subclusters and the further merging of such cluster in a supercluster (see Fig.~\ref{Local}).
We must underline that, as explained above, $\Psi_L(\tau)$ measures the Shannon information contained by a cluster related to the distribution of distances induced by diffusion evolution operator $e^{-\tau\hat L}$ in the cluster. In short, the higher the value of $\Psi_L(\tau)$, the more connected by diffusion is such cluster on the time-scale $\tau$. Figure \ref{Local}(a) shows a particular example of Zachary's karate club using the Laplacian, $\hat L$. As we illustrate in Fig. \ref{Local}(b), this module exhibits a large $\Psi_L(\tau)$ for all the range of significative timescales $\tau$ both for $\hat L$ and $\hat L_{RW}$ (see below).

\section{Temporal dendrograms and metastable nodes}
\label{metastable}

\new{As previously discussed, one central implication of the LRG framework is its {\em dynamic} perspective on the contribution of single nodes to diffusionally connected mesoscopic structures when the time-scales $\tau$ change. Mesoscopic divisions roughly maintain their internal composition (i.e., they are stable), but specific nodes can suddenly switch clusters at a certain $\tau$.
We focus on such nodes that, precisely for this property, act as a communication bridge between network regions. Therefore, we denominate this subset of nodes as {\em metastable} nodes, as they permit different highly connected subregions to share information. Consequently, metastable nodes can be attributed to different subdivisions at different resolution scales of the system. This explains why classifying these nodes is exceptionally challenging when using standard optimization algorithms that provide a static picture of the network. 
Due to the time-dependent definition of the distance matrix $\mathcal{D}_{ij}(\tau)$, it is straightforward to identify metastable nodes through the direct analysis of the cluster composition corresponding to the best partition of the dendrogram at increasing $\tau$.  }

\begin{figure}[hbtp]
    \centering
    \includegraphics[width=0.8\columnwidth]{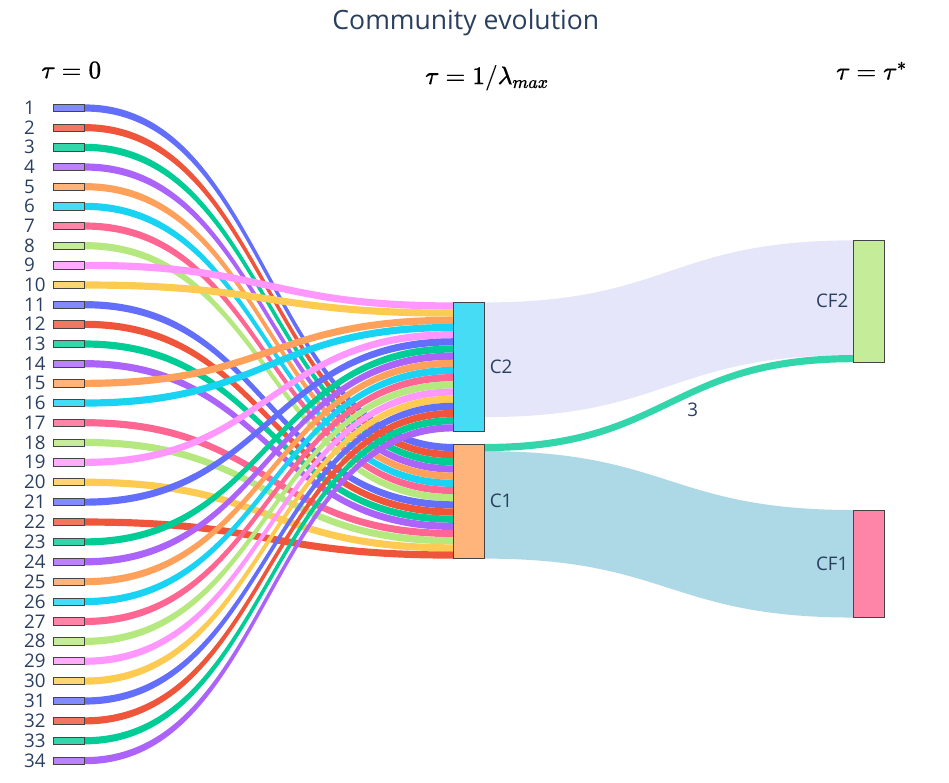}
    \caption{\textbf{Sankey diagram.} Evolution of the optimal community division (in terms of $\Psi$) for Zachary's Karate club at three different characteristic times. We emphasize the ability of 'bridge nodes' to dinamically change their functional community at different timescales. We estimate the characteristic time of change of Node 3 at $\tau=0.29(1)\simeq 5.26 \tau'$.}
    \label{Sankey}
\end{figure}

The Sankey diagram in Fig.~\ref{Sankey} illustrates this phenomenon, showing how Node 3 in Zachary's Karate club changes the module it belongs to at a specific $\tau$.
%whose community differs depending on the Laplacian we have previously considered (as discussed above). To shed light on this crucial fact, we now analyze this specific node by using $\hat L$. 
The figure clearly shows how the best assignment for this particular node changes at a specific time $\tau$ between $\tau'=1/\lambda_{max}$ and the peak of the specific heat at $\tau^*$, revealing its role as ``bridge" node. This feature assigns it a central role in communication and control processes between the two weakly dependent modules. In other cases, it is plausible to expect a similar relevance of metastable nodes in real networks (e.g., the human connectome) for synchronizability and information integration. The concept of metastable nodes generalize the one of ``modular flexibility" as defined for the dynamic correlation analysis of time-series in \cite{khambhati2018modeling} or previously described in community degeneracy problems \cite{ModFlex, Peixoto2021}.
The modular flexibility of a given node in the correlation network represents how frequently it changes the module it belongs to across time. This means that nodes are more likely to be connected to multiple modules at different times and are more flexible, being associated with network adaptability, and giving flexible nodes the role of pivoting the dynamics processes running on top of the network. 
%Although the concept of modular flexibility proposed by Khambati et al. is based on the description of the dynamic evolution of the correlation network of time-series, it can also be adopted and adapted to describe the modular properties of a network according to the dynamic regime of diffusive processes of information used to investigate it.

\begin{figure*}[hbtp]
    \centering
    \includegraphics[width=2.0\columnwidth]{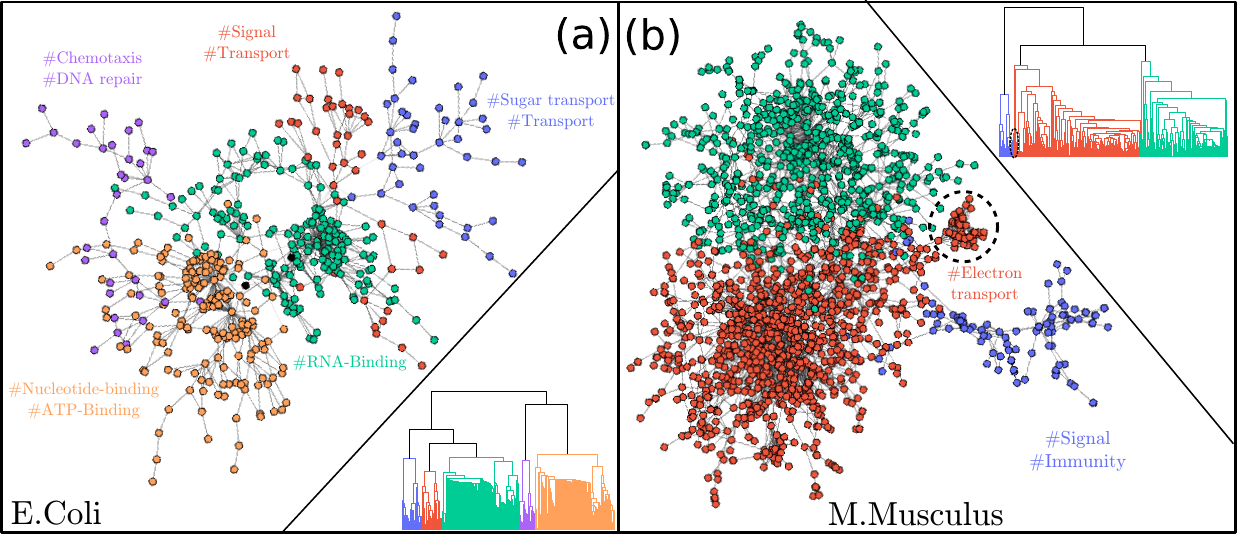}
    \caption{\textbf{Protein-protein interaction networks. (a)} Modular structure of the E.Coli PPI network for $\tau=10$. The dendrogram illustrates the hierarchical clustering of the different nodes of the network. Different colors represent different communities and their corresponding main biological functions. \textbf{(b)} Modular structure of the M.Musculus PPI network for $\tau=1$. The dendrogram illustrates the hierarchical clustering of the different nodes of the network. Different colors stand for different modules, as stated by the network dendrogram. Black nodes stand for metastable nodes. Note that instead of choosing the 'optimal' network partition in terms of communication distances, we are now interested in the complex nested structure and modules of both organisms, giving rise to a rich structure with multiple overlapping communities. The different functionalities associated with each module are written in the same color as the nodes it contains. We stress the characterization of a small local module responsible for electron transport as a direct application of $(\Psi_\ell)$ (see black dashed circle). Note that here, we have integrated out the microscopic scales of both networks without going so far as to integrate out the Fiedler eigenvector, ensuring a proper analysis of the network mesoscopic modules.}
    \label{Ecoli}
\end{figure*}

\new{
\section*{Disentangling PPI networks}
As important examples of real networks, we apply the full methodology described above (summarized in Appendix \ref{MSL}) to the complex structure of E. Coli and M. Musculus PPI networks \cite{Das2012}, as shown in Figure \ref{Ecoli}. We identify the functional role of these communities via the UniProt Knowledgebase \cite{UniProt}, a comprehensive, high-quality, and freely accessible set of around 190 million protein sequences annotated with functional information. In particular, we use the keywords associated with each protein to facilitate the characterization of different core functionalities of the observed modules. The PPI network of E.Coli presents a highly nested tree-like structure with heterogeneous submodules dedicated to other biological functions, as shown in Figure \ref{Ecoli}(a). For clarity, we explore the division into five modules exemplified in the dendrogram of Figure \ref{Ecoli}(a). In this case, two central communities emerge. The first one is related to RNA-binding (green nodes, ~88\% of the proteins engaged in this process belong to this community, constituting ~59\% of it), and the second one is inherent to chemotaxis and DNA repair (violet nodes, containing the 80\% of the proteins involved in chemotaxis, and conforming the 82\% of the community). We also identify two different particular metastable nodes (see black nodes in Fig. \ref{Ecoli}(a)):  'P0AGB6', a $\sigma^E$-factor responsible for the envelope stress response, and 'P0A6M8', an elongation factor G that catalyzes the GTP-dependent ribosomal translocation step during translation elongation.
Conversely, the M.Musculus network appears much more interconnected and highly intricate. Still, we can discern three principal modules, the smallest one of which (blue nodes in Figure \ref{Ecoli}(b)) is strongly related to the signal processing function and the immune response of the animal's cells. Finally, the direct application of $\Psi_L(\tau)$ allows the detection of a robust, independent local module (see black dashed lines in Fig. \ref{Ecoli}(b)), composed of a group of proteins dedicated to the Electron Transport Chain, a driver of the adenosine triphosphate (ATP) synthesis in mitochondria of eukaryotic cells. Even though a detailed study of different biological networks is beyond the scope of this manuscript, our results are highly illustrative of the power of the LRG, leaving extensive analyses of real structures for future works.}

\section{The Laplacian Random-Walk} 
\label{sLRW}

\begin{figure*}
    \centering
    \includegraphics[width=1.6\columnwidth]{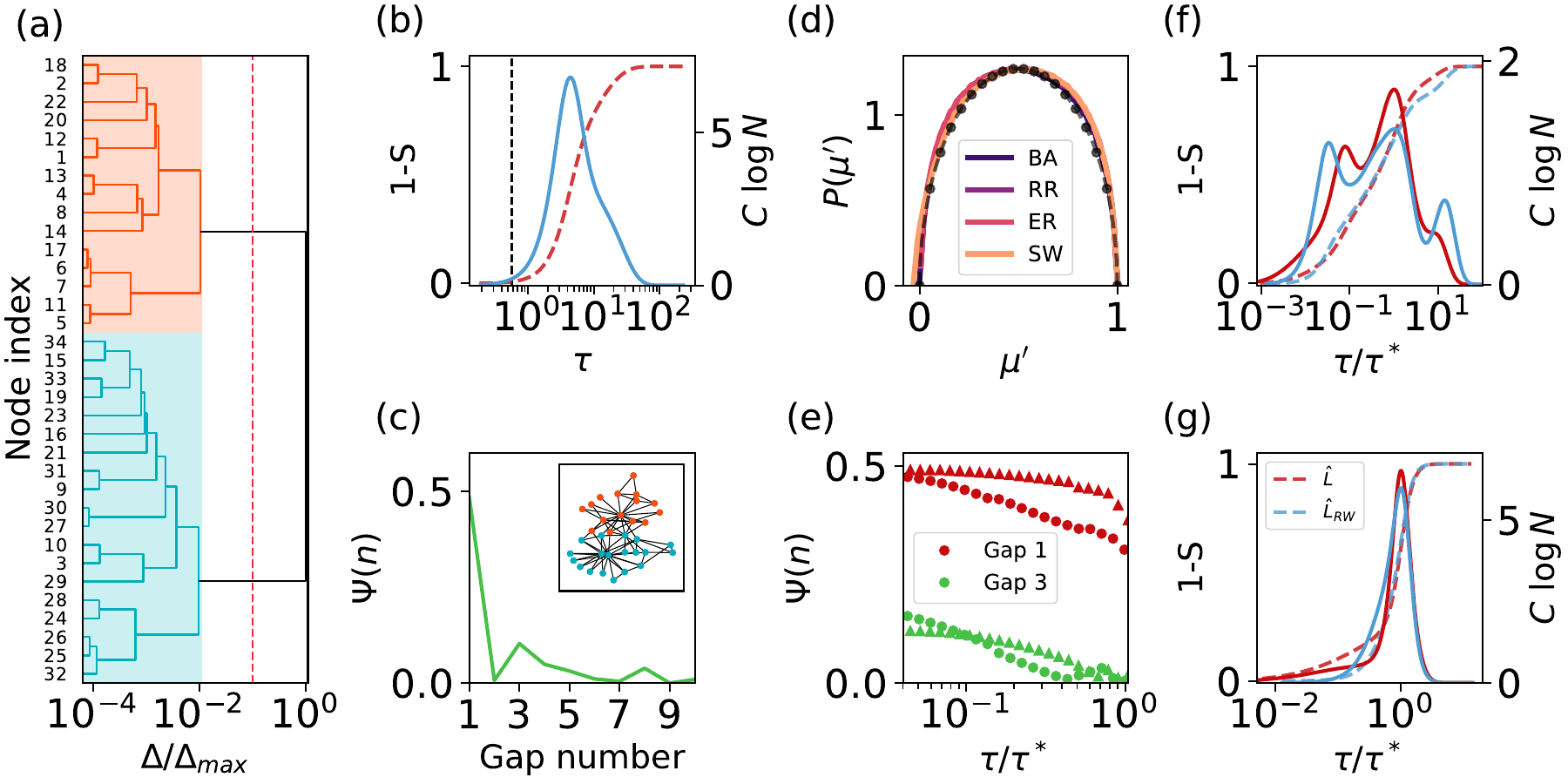}
    \caption{\textbf{Random-walk functional communities ($\hat L_{RW}$). (a)} Normalized dendrogram for Zachary's karate club with $\tau\lesssim\tau^\prime=/\mu_{max}$. Red dashed line reflects the optimal division as stated by $\Psi$. Different communities are shaded in different colors. \textbf{(b)} Entropy parameter (dashed red lines, $(1 - S)$), and specific heat (solid blue lines, $C$), versus the temporal resolution parameter of the network, $\tau$. Black dashed line marks $\tau'=1/\mu_{\max}$. \textbf{(c)} Partition Stability Index $(\Psi)$ versus gap number for $\tau=\tau^\prime$. Inset: Division into communities of the network. \textbf{(d)} Spectrum of eigenvalues for $\hat L_{RW}$ considering a Barabasi-Albert (BA) network with $m=10$, a Random Regular (RR) graph with $\kappa=18$, an Erd\H{o}s-R\'enyi (ER) network with $\langle \kappa \rangle=18$ and a Watts-Strogatz (SW) network with $\langle \kappa \rangle=18$ and rewiring probability $p=0.75$. Black dashed line stands for the semicircular law. \textbf{(e)} $\Psi$ versus $\tau/\tau^*$ (with $\tau^*$ corresponding to the maximum in C). Different colors represent different time gaps in the dendrograms for $\hat L$ (circles) and $\hat L_{RW}$ (triangles). \textbf{(f-g)} Evolution of the spectral entropy using the Laplacian and the Laplacian RW versus the normalized time, $\tau/\tau^*$ (where, for the sake of comparison, $\tau^*$ corresponds here to the absolute maximum of the specific heat, for (f) Mus Musculus PPI network and (g) Drosophila Melanogaster network. Note how eigenmodes are integrated out differently in both cases.}
    \label{RWalk}
    
\end{figure*}

One might ask whether the current use of the LRG can be extended to other meaningful dynamics on the network. To do so, we examine the stochastic behavior of random walks, using the random walk Laplacian $\hat L_{RW}=\hat D^{-1}\hat L$. This encodes a discrete time-traveling dynamics of a walker moving from a node to any nearest neighbor with uniform probability: i.e., the transition matrix for a RW on top of a graph \cite{Burioni2005,Chung1997} (see also Appendix \ref{LRWAp}). The operator $\hat\rho(\tau)$ is well-defined because, although $\hat L_{RW}$ is not symmetric, all its eigenvalues are real and fall within the range $0 \leq \mu_i \leq 2$. Notably, $\hat L_{RW}$ is similar to the symmetric operator $\hat L_{sym}=D^{1/2}\hat L_{RW}\hat D^{-1/2}$, enabling us to reformulate our framework for the RW exploration dynamics of the network by substituting $\hat L$ with $\hat L_{sym}$. 

In general, heterogeneous networks with sufficiently high connectivity are characterized by a spectral density that takes the well-known semicircular shape (Wigner semicircle law) in the interval $[2-\mu_{max}, \mu_{max}]$, which is centered at $\mu=1$ \cite{Chung2003, Dorogovtsev2022, Sakumoto2021, PeixotoSpec}. This suggests a common universal dynamic among these diverse structures. To better analyze the spectrum of $\hat L_{RW}$, the eigenvalues spectrum can be standardized to collapse the semicircular (Gaussian) part of the spectrum. A perfect collapse is expected for networks with a large enough lowest degree \cite{Mendes2008} (roughly corresponding to a sufficiently large effective value of the spectral dimension, $d_S$, see Appendix \ref{LRWAp}). The standardization consists of the simple transformation $\mu'=\frac{1}{2}+\frac{\mu-1}{2(\mu_{\max}-1)}$. Instead, the negative part of the spectrum in the standardized eigenvalues $\mu'$ (except the trivial lowest eigenvalue $\mu=0$ corresponding to $\mu'=\frac{\mu_{max}-2}{2(\mu_{\max}-1)}<0$) becomes absent for completely random cases, while it is expected when rare regions effects --often leading to anomalous slow relaxation of the system-- are present \cite{Juhasz2012} (see Appendix \ref{LRWAp}). Fig. \ref{RWalk}(d) reports different benchmark monopartite networks where the spectrum of $\hat L_{RW}$ converges to this expected universal shape \footnote{Indeed, as the density matrix encodes pairwise information exchange between nodes, we expect the existence of such universal behavior for an effective spectral dimension $d_S^*=4$, similarly to the Gaussian model \cite{Donetti2002,Kardar,Cassi1999,Duplantier1988}}.

% This is because the random walk Laplacian spectrum of the network is closely related to the first-return times and the distribution of all returns induced by the network \cite{Mendes2008,Mitrovic2009} 
%(but not the fluid Laplacian, $\hat L$, where the eigenvectors form the Fourier basis of the network where the random walk dynamics is projected through $\hat L_{RW}$).

Figure \ref{RWalk}(a) shows the RW exploration dendrogram for Zachary's karate club. We highlight that, for $\tau$ equal to the inverse of the maximum eigenvalue, Node 3 belongs to a different community using $\hat L$ or $\hat L_{sym}$ (comparing with Fig. \ref{Zach}(c)), indirectly confirming its metastable nature. Figure \ref{RWalk}(b) shows the entropy $S_{RW}(\tau)$ calculated now using the RW density matrix $\hat \rho_{RW}(\tau)=\exp(-\tau\hat L_{sym})/Z_{RW}(\tau)$, and the related susceptibility (i.e. specific heat) $C_{RW}(\tau)$ for the same network. Note that $C_{RW}$ presents a slightly different shape compared to the one of Fig. \ref{Zach}(b) obtained by the Laplacian $\hat L$, as an effect of the dynamical modes of the network induced by $\hat L_{sym}$ (i.e. $\hat L_{RW}$). Figure \ref{RWalk}(c) compares the particular time $\tau^{\prime}=1/\mu_{max}$ determined by the maximization of $\Psi(n;\tau)$ (see Fig. \ref{RWalk}(e) for an analysis of the first two optimal gaps over multiple timescales).
Figure \ref{Local}(b) shows the comparison of the behavior of the Local Partition Stability Index $\Psi_L(\tau)$ for the same branch of the dendrograms at different times using both $\hat L$ and $\hat L_{sym}$. In addition, we have also characterized the characteristic change time for metastable Node 3, $\tau=1.94(1)\simeq 3.31 \tau'$.

Two questions naturally arise at this point: (i) how different are the partitions obtained by the heat/diffusion equation Laplacian $\hat L$ \pv{concerning} those obtained by the symmetrized version of the random walk Laplacian $\hat L_{RW}$? (ii) Is there some {\em physical} criterion to decide when \pv{it} is better to use the former or the latter?

Figures \ref{RWalk}(f) and (g) show the comparison of the two Laplacian entropies and specific heats (or entropic susceptibilities) for the \emph{Mus Musculus} PPI network \cite{Das2012} and the color vision circuit in the medulla of \emph{Drosophila Melanogaster} \cite{Takemura2013}. They show how the two Laplacians integrate information differently in the two networks. In particular, these effects will strongly depend on the effective (spectral) dimension of the network, both at global or local scales \cite{Pruessner2019}. As a simple and illustrative example of this effect, one can consider the analysis of the Star-Clique network (see Appendix \ref{CST}): a clique of $m\gg 1$ nodes is connected to a star of $n\gg 1$ nodes by a single bridge edge. A random walker will be effectively trapped at the star hub or in the clique for a long time before visiting both structures. This leads to two well-defined and dynamically separated communities using $\hat L_{RW}$. Instead, the diffusion flow with a uniform rate per edge, described by the Laplacian $\hat L$, which additionally captures the broadcast of information linked to both hubs, giving place to more accurate information on the topological properties of the network (see Appendix \ref{CST}).

The major difference between $\hat L$ and $\hat L_{RW}$ is that $\hat L$ is extensive in the local connectivities of the network while $\hat L_{RW}$ is intensive, and the stochastic flow is normalized over the entire neighborhood of every single node. This gives rise to trapping mechanisms in the RW when two highly connected subregions communicate weakly. When choosing which of the two operators is more appropriate, another aspect that must be considered is their relation to eventual dynamical models on the network to be studied. Statistical and dynamical models on graphs, such as epidemic and contact processes, Ising and Kuramoto models, when written in terms of continuous fields on the nodes of the graph, present a two-body interaction term, giving the Gaussian approximation, substantially determined by $\hat L$. This is consistent with previous studies underlining the relevance of using $\hat L$ instead of $\L_{RW}$ for community detection purposes \cite{Lambiotte2014}. This is because if the strength of the interaction between any pairs of neighboring nodes is fixed, then the global interaction that a single node has with the rest of the network is extensive in the local connectivity, i.e., the node degree. This is the case of the Laplacian $\hat L$, while in the operator $\hat L_{RW}$, it is the global interaction that any single node has with the rest of the network to be fixed to a unitary constant. For this reason, when one is interested in the best network partition for this kind of systems, the natural choice is to use $\hat L$.

%$$\hat L$ contains the genuine, plain topological information of the network, while $\hat L_{RW}$....

\section{Discussion and conclusion}

The strong topological heterogeneity constitutes a fundamental issue in real networks \cite{Garcia2018}, inducing non-unique and highly complex algorithms for partitioning and community detection. Grounding on the intrinsic property of the Laplacian time-evolution operator $e^{-\tau\hat L}$ to be a sort of telescopic "scanner" of the mesoscopic scales of a network, we propose a simple but efficient application of the Laplacian Renormalization Group theory \cite{InfoCore, LRG}, which permits the introduction of an ultrametric communication distance related to the diffusion of information to disentangle the mesoscopic structure of any heterogeneous system. This naturally offers a natural and parsimonious interpretation of the so-called {\em resolution limit} \cite{Fortunato2022,Dorogovtsev2022} that generally affects modularity community detection algorithms:  it has been shown that using the modularity function is equivalent to limiting the weighting scheme for the diffusional paths to the linear contribution in the Laplacian operator. On the contrary, weighting the different paths from vertex $i$ to vertex $j$ through the operator $e^{-\tau\hat L}$  implies properly considering the contributions coming from all the significant powers of $\hat L$ for each value of the scale parameter $\tau$. Hence, by varying $\tau$ from the minimal significant value $\tau'=1/\lambda_{max}$ and the maximal one $\tau^*$, corresponding to the large time peak of $C(\tau)$, we can resolve all the telescopic community structure of the network. This explains the \emph{defocusing} of small-scale structures by increasing $\tau$. Once a certain number of diffusive eigenmodes has been integrated out, it results in a loss of information about the microscopic network scales related to that part of the spectrum. 

Our framework has important theoretical implications. Firstly, it establishes a substantial (but not rigorous) equivalence between real-space methods derived from the $\hat \rho(\tau)$-density matrix (i.e., Louvain methods for community detection or Markov stability) and spectral clustering techniques \cite{Fiedler1973,Pothen1990,Donetti2004}, which addresses the issue in $k-space$ and analyzes diffusion modes of the network. These methods can be seen as two sides of the same coin. Furthermore, it provides a unified interpretation of community detection in terms of structural and dynamic aspects, offering an overall vision of different methodologies that were previously considered separately.

We underline that the interaction matrix of the Gaussian field theory on graphs \cite{Donetti2002, Erzan2015} is always defined in terms of the Laplacian $\hat L$ when the global interaction that a single node has with the rest of the network is extensive in the local connectivity. This is the basis of the Gaussian approximation of any statistical model on graphs (Ising, Kuramoto, contact-epidemic models, etc). As a result, we stress the general importance of using $\hat L$ (not $\hat L_{RW}$) for community detection purposes on top of matrices resulting from dynamical processes in which the interaction of the single nodes with the rest of the networks is the sum of the uniform interactions with all neighbors. The use of $\hat L_{RW}$ has to be limited to the RW and related stochastic models for the diffusion of single particles on the network substrate.

As a direct consequence, the eigenvalues spectrum of $\hat L_{RW}$ takes the universal semicircular shape \cite{Chung2003}, independently of the peculiar topology of the network, when the minimal connectivity is sufficiently large \cite{Mendes2008}. This quantity has been shown to control the emergent functional communities and dynamical aspects of the system \cite{Cassi1999}. We emphasize that the spectrum does not provide any local topological information on the heterogeneous network, as long as the mean and minimal connectivity are finite and sufficiently large: a single RW will visit a negligible fraction of the network in the large time limit. This means that different realizations of the process may yield completely different information on the local topology of the network, since their intersection is negligible. A similar point has been addressed in \cite{Duplantier1988}, in the context of Euclidean spaces, where a strict relation with the upper critical dimension of $\phi^4$ field theory is established. This topological limitation does not affect the Laplacian $\hat L$, which is always sensitive to local topological features. The reason grounds on its extensivity in the local connectivity of nodes. Indeed, independently of the minimal connectivity of the heterogeneous networks, the {\em fluid} Laplacian $\hat L$ \cite{Masuda2017}, through its time-evolution operator $e^{\tau\hat L}$, describes a process that gradually invades all the network.

%It is still important to discuss the subtle link between the Gaussian model and the random walk on any graph (we also refer to \cite{Donetti2002,Burioni2005} for an extended discussion on the issue). In particular, $d^*_S=4$ characterizes a critical value for the intersection properties of two independent random walks, which is closely connected to the non-triviality of the $\phi^4_d$ field theory \cite{Duplantier1988}. Note that over this dimension, two random walkers are unlikely to intersect, and the Gaussian results obtained neglecting the intersections are asymptotically valid \cite{Kardar}), thereby, making the spectra of the $\hat L_{RW}$ to show some type of universal behavior \cite{Chung2003}. 
%We want to emphasize that the lowest vertex degree in a network will critically constrain its spectral dimension \cite{Mendes2008}, which control the emergent functional communities and dynamical aspects of the system \cite{Cassi1999}. 
%We recognize the relevance of this measure intending to discriminate the relevance of underlying heterogeneity on the system dynamics (e.g., scale-free networks are expected to be dynamically irrelevant for large enough values of $m$ in the Barabasi-Albert model \cite{Mendes2008}).

We have evidenced a further crucial feature of heterogeneous networks: the existence of a subset of {\em metastable} nodes that, depending on the scale of observation of the network, can switch the community to which they belong. Understanding the general features of this class of nodes can be particularly relevant for controllability aims. For instance, one can expect this switching behavior on different scales $\tau$ to play a central role in forming attractors as expected for synchronization dynamics in hierarchic modular networks \cite{Villegas2014,Villegas2016}.
A similar central role may be played in the metastable dynamics in human brain networks, where hubs have been elucidated to control the resulting wave patterns \cite{Gollo2019}. We conjecture that this subset of nodes may be involved in the observed disrupted hub organization in the topological disturbances associated with schizophrenia \cite{Fornito2014}, where the waste of hubs \cite{Bullmore2015} can lead to dysfunctions in controlling and integrating neuronal signals \cite{Klauser2017}. %Finally, our new multi-scale Laplacian (MSL) approach sheds light on the origin of the resolution limit problem while giving a global overview of the importance of local communities in the entire network \cite{Fortunato2022}.

We are aware of the computational burden represented by computing the full exponential of a matrix when analyzing the intrinsic network structure. Still, we offer a complete solution to scrutinize the full range of scales of a network by considering its entire set of eigenvalues. Instead, when dealing with large networks, we suggest considering a projection of the most significant eigenvalues of the N-dimensional k-space of $\hat L$ using the components of their eigenvectors as coordinates (a natural extension of previous spectral methods \cite{Donetti2004}, absent of any null-model). In a nutshell, this is the Taylor expansion of the Laplacian matrix up to the n-th order. Much further work is still needed to better understand the relevant number of eigenvalues as a function of the nature of the network.

Altogether, we propose here a way of disentangling the intertwined mesh structure characterizing complex networks at every scale in the light of the recently formulated LRG \cite{LRG}. Also, the proposed formulation does not need any null-model to detect significant modules and naturally identifies the ``building" blocks of the network at different scales, controlling for resolution limit constraints. Moreover, we shed light on the very origin of the resolution limit problem while giving a global overview of the importance of local communities in the entire network \cite{Fortunato2022}. %In particular, we scrutinize the direct links with previous definitions of network modularity, present an overall picture of different frameworks, and reveal their limitations. 

Our approach opens a new perspective on the problem of optimal partition in communities of heterogeneous networks, as grounds on a fundamental dynamical operator that plays a central role in many statistical models of interactions on networks. Such can be generalized to other dynamical processes on networks \cite{Rosvall2014,ghavasieh2022}, and related operators if one is interested in the optimal partition corresponding to these dynamics to predict functional communities.   

\begin{acknowledgments}
P.V. acknowledges the Spanish Ministry and Agencia Estatal de investigación (AEI) through Project I+D+i Ref. PID2020-113681GB-I00, financed by MICIN/AEI/10.13039/501100011033 and FEDER ‘A way to make Europe’. We also thank G. Cimini, F. Saracco, G. Caldarelli and D. Garlaschelli for useful suggestions and comments.
\end{acknowledgments}

\appendix

\section{Statistical physics of information network diffusion}
\label{StatPhys}

Let $\hat L$ be the Laplacian associated with an undirected monopartite network, namely $
L_{i j}=\left[\left(\delta_{i j} \sum_k A_{i k}\right)-A_{i j}\right]$,
which governs the graph Laplace/heat equation,
\[ 
\dot{\textbf{s}}(\tau)= - \hat L \textbf{s}(\tau)\,,
\]
we have specifically considered its time evolution operator $e^{-\tau\hat L}$, which is on the basis of a canonical description of heterogeneous networks and encodes the topological properties of the network. Given a probability distribution $s(\tau=0)$, its temporal evolution is given by $s(\tau)= e^{-\tau \hat L}s(\tau=0)$. 
In that discrete-states representation, each element of the propagator $e^{-\tau \hat L}$ describes
the sum of diffusion trajectories along all possible paths
connecting nodes $i$ and $j$ at time $\tau$ \cite{Masuda2017, Burioni2005, Moretti2019}.
Normalizing the propagator, it is possible to define the ensemble of accessible information diffusion states \cite{InfoCore, Domenico2016, Ghavasieh2020}, obtaining
\begin{equation}
\hat{\rho}(\tau)=\frac{e^{-\tau \hat{L}}}{\operatorname{Tr}\left(e^{-\tau \hat{L}}\right)},
\label{r}
\end{equation}
where one can recognize in $\hat{\rho}(\tau)$ the form of a canonical density operator \cite{binney1992theory, greiner2012thermodynamics, pathria2011edition}. Note that, in full analogy with Hamiltonian systems in statistical physics, $\tau$ and $\hat L$ play the role of $\beta$ and ${\cal H}$, i.e. the inverse temperature and the Hamiltonian function, respectively. It is important to stress that we consider connected networks to fulfill the ergodic hypothesis. At that point, one can define the network entropy \cite{Domenico2016} as
\begin{equation}
\begin{aligned}
S[\hat{\rho}(\tau)]&=-\operatorname{Tr}[\hat{\rho}(\tau) \log \hat{\rho}(\tau)]\\
&=-\frac{1}{\log N} \sum_{i=1}^N \nu_i(\tau) \log \nu_i(\tau),\end{aligned}
\end{equation}
where $\nu_i$ represents the set of eigenvalues of $\hat \rho$.
Through the detailed analysis of the entropy flux, it is possible to track the entropy-driven transition over the network \cite{InfoCore}. In particular, this passes from a strict fragmentation at $\tau = 0$, where $S=0$ and the system lies in a segregated phase, to a uniformly connected through diffusion graph, where $S=1$ and the system lies in an integrated and homogeneous phase. The derivative of the entropy of the logarithm of the diffusion time $\tau$,
% The asymptotic homogeneous phase is given by the symmetry of the markovian operator $\mathds{1}-L$, as it is possible to write $s(t+1)=(\mathds{1}-L)s(t)$ in the discrete- time approximation.
\begin{equation}
C(\tau)=-\frac{d S}{d \log \tau}
\label{cvv}
\end{equation}
is a detector of structural transition points corresponding to the intrinsic characteristic diffusion scales of the network \cite{InfoCore,LRG}. Indeed, a pronounced peak of C defines $\tau=\tau^*$ and reveals the starting point of a strong deceleration of the information diffusion, separating regions sharing a rather homogeneous distribution of information from the rest of the network. If more well-separated diffusion timescales exist, then $C(\tau)$ can show a multi-peak structure.

\section{Laplacian Renormalization Group}
\label{LRGAp}

The renormalization problem is approached here \textit{\`a la Wilson} (we refer to \cite{LRG} for a full discussion on the issue, in particular, concerning the real-space procedure), carrying on the comparison with the canonical ensemble, as shown in Appendix \ref{StatPhys}. The first step consists of moving to the Fourier space to analyze the network eigenmodes (as the graph lacks of any spatial embedding). One may anyway keep in mind that $\hat L$ contains the inverse of the diffusion time scales. As expected from a discrete version of Gaussian dynamics in the continuum $\kappa$-space, $\hat L$ is diagonalizable, and the change of basis leads to a decoupling of modes. Since $\hat L$ is symmetric and real-valued, it holds a complete set of eigenvectors $\{|\lambda\rangle\}$, with semi-positive eigenvalues $\{\lambda\}$.
\\In the bra-ket notation, the Laplacian operator can be decomposed as the projector  $\sum_\lambda \lambda|\lambda\rangle\langle\lambda|$. The LRG step consists of integrating out these diffusion
eigenmodes from the Laplacian and appropriately rescaling the network, namely:
\begin{enumerate}
\item \label{un}Reduce the Laplacian operator to the contribution of the $N-n$ slow eigenvectors with $\lambda<\tilde{\lambda}, \tilde{\hat L}=\sum_{\lambda<\tilde{\lambda}} \lambda|\lambda\rangle\langle\lambda|$;
\item \label{du} Then rescale the time $\tau \rightarrow \tau^{\prime \prime}$, so that $\tilde{\tau}$ in $\tau$ becomes the unitary interval in the rescaled time variable $\tau^{\prime \prime}: \tau^{\prime \prime}=\tau / \tilde{\tau}$ and, consequently, redefining the coarse-grained Laplacian as $\hat{L}^{\prime \prime}= \tilde{\tau} \tilde{\hat{L}}$. 
\end{enumerate}
As for a RG procedure applied to a Gaussian system, one could expect to recover the same original diagonal form after step \ref{du}. Concerning the first step, point \ref{un} can be implemented by letting the time run from $\tau'=1/\lambda_{max}$ to a value $\tau^*\sim 1/\lambda_{gap}$. As a direct consequence of this, looking at the propagator $e^{-\tau \hat L}$, it is possible to observe that the contribution to the measure of a given eigenvalue $\lambda$ starts decreasing significantly when $\tau \sim 1/\lambda$. Such soft pruning overcomes the difficulties introduced by the non-euclidean support, integrating the different network eigenmodes one by one and saturating, as $\tau$ increases, the smaller system scales. In particular, note that this effect is automatically encoded in the $\hat\rho-matrix$ that we have defined in the main text to define the \emph{Communication distance} between nodes. Now the first step involves a specific cut in the dendrogram.

%moreover, since the set of eigenvalues is not dense in finite-size networks, a such soft cut may anyway appear strict, if a sufficient gap in the eigenvalues set occurs in the shell boundary. Such request is well satisfied where the specific heat in Eq. \ref{cvv} shows a peak.
\pv{
\section{Community detection in the light of the LRG} \label{CDetection}
Formulating a unifying framework that can help us understand community detection from first physical principles is a crucial challenge partially addressed by the concept of Markov stability \cite{Barahona2010, Lambiotte2014}.

It compares, for each arbitrary partition $\{g_i\}_{1=1}^N$ of the network in communities, for all pairs of nodes belonging to the same community, the joint probability that a discrete-time random walker at the stationary state is at node $i$ at time $t=0$ and node $j$ at time $t=\tau$, with the null model given by the same joint probability in the asymptotic limit $\tau=+\infty$. Hence, if one considers the particular choice where the waiting times of a discrete-time random walk are inversely proportional to its degree --i.e., an inhomogeneous rescaling of the time-- the combinatorial continuous-time Markov stability reads \cite{Lambiotte2014, Barahona2010},
\begin{equation}
R(\tau)=\stackrel[i,j=1]{N}{\sum}\left[p_{i}^{*}\left(e^{-\tau \hat L/\langle \kappa\rangle}\right)_{ij}-p_{i}^{*}p_{j}^{*}\right]\delta\left(g_{i},g_{j}\right),\label{MSTAp}
\end{equation}  

where $\hat L=\hat D- \hat A$ is the ordinary Laplacian matrix of a network,  $\tau$ is a time resolution parameter, which allows the unraveling of multiscale structures in the network considering diffusion at all different values of $\tau$, $\langle \kappa \rangle$ is the average degree of the network and the null probability $p_{i}^{*}p_{j}^{*}$ corresponds to the expected transfer probabilities at stationarity for this Markov process \cite{Lambiotte2014}.

It is important to underline that, for an undirected network with $N$ nodes and $M$ edges, the modularity of a given partition (see \cite{Fortunato2010} and references therein), 
\begin{equation}
Q=\frac{1}{2M}\stackrel[i,j=1]{N}{\sum}\left(A_{ij}-\gamma P_{ij}\right)\delta\left(g_{i},g_{j}\right)\label{ModulAp},
\end{equation}
can be easily derived from Eq\eqref{MSTAp} by Taylor-expanding the exponential $e^{-\tau \hat L}\simeq \hat I-\tau \hat L$ \cite{Masuda2017}. Here, $\hat A$ stands for the adjacency matrix of the graph, $\gamma$ is a resolution parameter, and $g_i$ is the index of the community to which node $i$ belongs, being $\delta(g_i,g_j)=1$ if $g_i=g_j$ and $0$ otherwise. $P_{ij}$  is the adjacency matrix of a suitable null-model that critically drives the modularity maximization process over all possible partitions in communities \cite{Garlasca2015}. In this particular case, it has been demonstrated that $P_{ij}$  is the adjacency matrix of the Erd\H{o}s-R\'enyi null-model for the combinatorial Laplacian. However, \pv{even if Markov stability still needs a null model to maximize $R(\tau)$,} Eq.\eqref{MSTAp} provides a simple interpretation of the resolution parameter ($\gamma=\nicefrac{1}{\tau}$), in terms of the evolution time of the Markov process \cite{Masuda2017,Barahona2010}.

Let us briefly discuss the scenario where $\hat L_{RW}=D^{-1} \hat L$, the \emph{asymmetric normalized Laplacian} or \emph{random-walk Laplacian}, is plugged into Eq.\eqref{MSTAp}. This involves considering the configuration model as the null model in the 'usual' modularity \cite{Lambiotte2014}. Note that the same approximation leading from Eq.\eqref{MSTAp} to Eq.\eqref{ModulAp} can be applied in this case. In addition, we refer to \cite{Barahona2010} and \cite{Lambiotte2014} for further connections with other community detection algorithms based on global optimization methods.

Null models have played a crucial role in defining quality functions for community detection algorithms. However, their accuracy is still a matter of debate \cite{Garlasca2015}, and there is a strong relationship between the modules detected in a network and the null models used to detect them. Instead, the precise analysis of the network propagator ($e^{-\tau \hat L}$), which constitutes the core of Eq.\eqref{MSTAp} and, more importantly, the formulation of a Laplacian-based statistical physics framework for heterogeneous structures, which tightly couples temporal and spatial scales \cite{LRG, InfoCore}, is essential to overcome this conundrum.

Equation Eq.\eqref{r} provides a clear understanding of the distances between network nodes that cannot be embedded in regular Euclidian spaces. It interprets the node distances in terms of network eigenmodes, which gives a natural interpretation \pv{of} spectral methods. This equation extends the algorithm proposed in \cite{Donetti2004}, but considering all possible eigenvectors in the N-dimensional space of the network. Note that $\mathcal{\hat D}$ is a projection of how points (nodes) are distributed in the N-dimensional k-space representation of the network. We stress that the required number of eigenvalues and eigenvectors to capture the intrinsic structure of a network depends on its complexity and heterogeneity and, therefore, further research is still needed in this direction. In summary, our framework connects real-space and k-space insights \cite{LRG} by considering all available eigenvalues --and consequently the complete structural information-- of the network.
}

\new{\section{Identyfing mesoscales through the LRG}
\label{MSL}
The practical application for the identification of specific mesoscopic structures through the LRG can be summarized in the following fundamental steps: 
\begin{enumerate}
    \item For fixed $1/\lambda_{max}\simeq \tau\lesssim 1/\lambda_{gap}$, compute  
     the communication distance matrix $\mathcal{D}_{ij}(\tau)=\frac{(1-\delta_{ij})}{K_{ij}(\tau)}$, where $\hat K (\tau)=e^{-\tau\hat L}$, which permits to define an ultrametric distance between each pair of nodes on the diffusion time scale $\tau$.
    \item  Apply a usual algorithm to solve clustering problems in a metric space, such as, for instance, the average linkage clustering, to define a dendrogram. An arbitrary cut at a distance threshold $\Delta$ of such a dendrogram will give a particular network subdivision on this time scale.
    \item In order to select the optimal cut of the dendrogram, one has to detect the threshold value $\Delta$ corresponding to the maximum value in $n$ of the Partition Stability Index $\Psi(n;\tau)$ defined in Eq.~\eqref{psi-n}.
    \item The diffusion time-scale is arbitrary, but the analysis of the network Laplacian entropy, $S(\tau)$, and more specifically of the entropic susceptibility, $C(\tau)$, can be used to detect the characteristic mesoscopic scales of the system and select particularly significant values of the resolution parameter $\tau$. For instance, by selecting $\tau$ in the region just before the first peak, all the microscopic details of the system are erased, however, conserving all the large intrinsic characteristic structures related to diffusion. 
    \item An important criterion for the significance of the best partition at a given $\tau$ is its stability across a sufficiently large window of time-scales except for a few metastable nodes (see previous section) that act as intercluster bridges switching clusters by varying $\tau$.
\end{enumerate}}

\section{Unraveling nested structures in heterogeneous networks}

\pv{\subsection*{Stochastic-block models}

The SBM, or stochastic block model, is one of the statistical paradigms in network theory due to its ability to distill complex modular architectures into a simple structure. In essence, the SBM entails dividing the nodes of a network into $K<n$ distinct communities or groups. In particular, there are $K$ groups of $N$ nodes, and each node $i$ has a label $l_i\in\{1,...,K\}$. The edges connecting the nodes are generated based on the internal connectivity of each group, denoted by $\kappa_{in}$, and the external connectivity between the different groups, denoted by $\kappa_{out}$. It is worth noting that when $\kappa_{in}$ is equal to $\kappa_{out}$, the network is formally equivalent to an Erd\H{o}s-R\'enyi network.

\begin{figure}[hbtp]
    \centering
    \includegraphics[width=1.0\columnwidth]{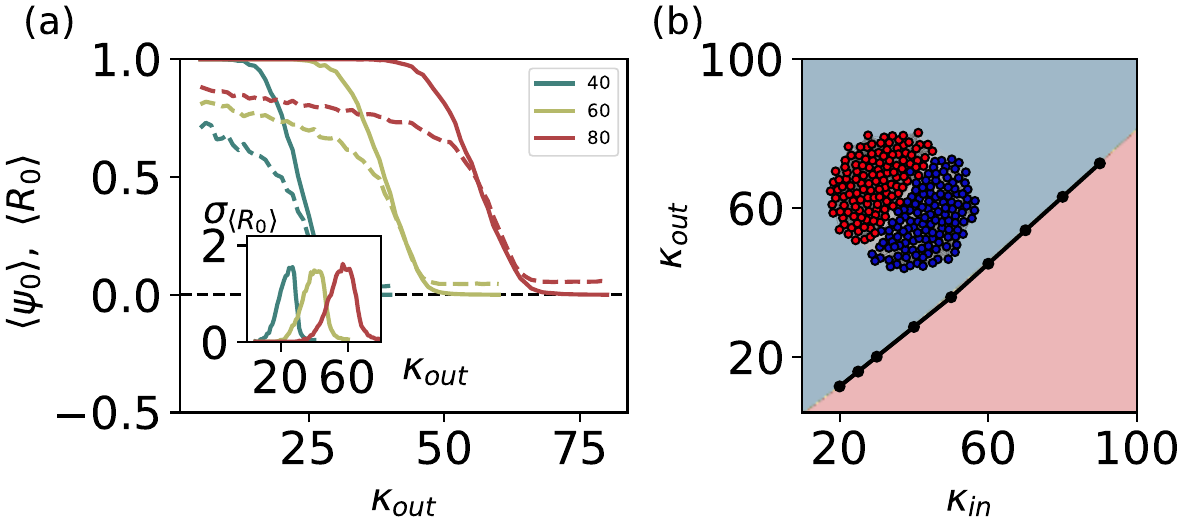}
    \caption{\textbf{Stochastic block model. (a)} Averaged Maximum Partition Stability Index, ($\langle \Psi_0 \rangle$, dashed lines), and Adjustes Rank Index ($R_0$) as a function of the external connectivity, $k_{out}$. Different colors represent different choices of the inter-block connectivity ($k_{in}$, see legend). Inset: Fluctuations of  the Adjusted Rank Index, $\sigma_{R_0}$, as a function of the external connectivity, $k_{out}$. The peak of the fluctuations acts as a detector of the detectability phase transition for the two-block model. Parameters: $N=256$. \textbf{(b)} Phase diagram for the detectability region of the two-block model. Black solid line represents the peak of $\sigma_{R_0}$ for different values of $k_{in}$.}
    \label{SBM2}
\end{figure}

In particular, for the simplest possible case with $K=2$ groups of equal size, $\nicefrac{N}{2}$, it has been demonstrated that the block model loses any community structure when at the specific condition \cite{Newman2012, Moore2014}, namely,
\begin{equation}
    \kappa_{in}-\kappa_{out}=\sqrt{2(k_{in}+k_{out})}.
    \label{Detect}
\end{equation}

In the case of multiple groups, a detectability transition can be established by generalizing this condition (we refer to \cite{Moore2014} for an extended discussion on the issue). Here, we focus on two communities of equal size ($K=2$), for which the detectability region, as defined by Eq.\eqref{Detect}, is highlighted in pink color in Figure \ref{SBM2}(b). To characterize the quality of the partition, we perform the following analysis: (i) we select the time at which the specific heat reaches its maximum, denoted by $\tau^*$, to construct the network dendrogram (using the Ward's method \footnote{Note that the selection of linkage depends majorly on how points are distributed in the N-dimensional k-space representation of the network.}) as stated in the main text. (ii) We identify the particular node division at the maximum value of the Partition Stability Index, denoted by $\Psi(n)$, which we refer to as $\Psi_0$. 

Figure \ref{SBM2}(a) shows the evolution of $\Psi_0$ versus the external connectivity between the different groups, denoted by $\kappa_{out}$, making transparent the high modular structure of the network. In particular, $\Psi_0$ goes to zero just at the detectability transition predicted by Eq.\eqref{Detect}. Indeed, we compare the obtained separation in modules with the true one employing the Adjusted Rank Index, $R_0$,  which quantifies the agreement in the ordinal placement of elements while accounting for the potential agreement by chance \cite{RankIndex,RI2}. Let us emphasize the perfect agreement of the detectability phase transition predicted by the MSL framework and the theoretical expectation of Eq.\eqref{Detect}.}

\begin{figure}[hbtp]
    \centering
    \includegraphics[width=1.0\columnwidth]{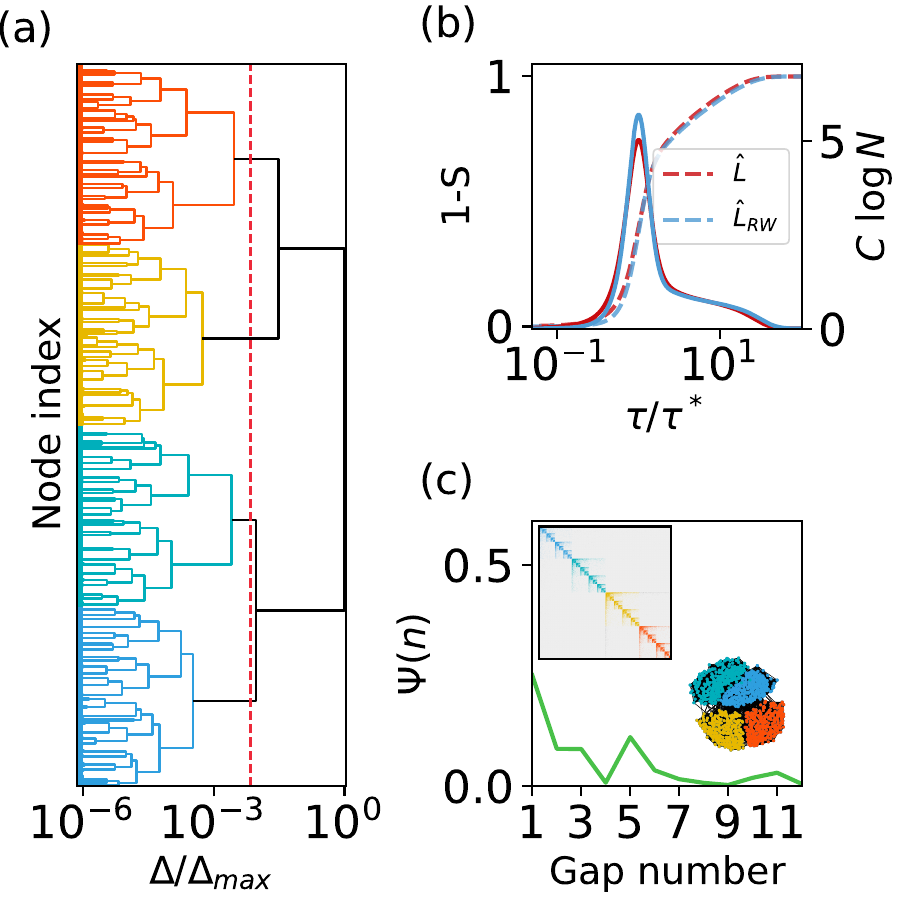}
    \caption{\textbf{Hierarchic modular network. (a)} Normalized dendrogram for a HM-CP network using $\tau'=1/\lambda_{\max}$. Red dashed line reflects the division of the network using the third gap of $\Psi$. Different communities are shaded in different colors. \textbf{(b)} Entropy parameter (dashed lines, $(1-S)$), and specific heat (solid lines, $C$), versus the temporal resolution parameter of the network, $\tau$. \textbf{(c)} Partition Stability Index $(\Psi)$ versus gap number for $\tau=\tau^\prime$. Note the high values of $\Psi$ indicating the hierarchical structure of the network. Insets: Adjacency matrix and division into four communities of the network.}
    \label{HM-CP}
\end{figure}

\subsection*{Hierarchic modular networks} \label{HMCP}
\pv{Let us discuss an even more complicated case than a stochastic block model with K communities or the random nested case with scale-dependent probability, both of which are unproblematic for the MSL framework.}
We consider a specific case of hierarchical networks where the connections between modules are not left at random but depend on a scale-dependent probability, promoting centralized structures between hubs and following the algorithm proposed in \cite{Zamora2016}. We thus create at the beginning $2^s$ blocks of $N_s=16$ nodes with mean degree $\kappa_0=12$ at the deepest level. Once this has occurred, we give a weight $p(i)=\nicefrac{i^{-\alpha}}{\sum_j j^{-\alpha}}$, to the $i^{th}$ node of each block, $i=1,2,...,N_s$. At this point, nodes are selected with probability $p(i)$ and $p(j)$, and connected if they were not already linked.  We use here the same scale-free exponent $\alpha=2$ for all the hierarchical levels except for the basal one, with $\alpha=1.7$ (as suggested to \pv{mimic} the empirically supported core-periphery organization with connector hubs in brain structural networks \cite{Eguiluz, Bassett-core, Bassett-core2}).

Figure \ref{HM-CP}(a) illustrates the dendrogram for a specific network and the nested nature of the different modules, which present the expected aggregation in the different hierarchical scales. We also present here the comparison of the spectral entropy of the network by using $\hat L$ and $\hat L_{RW}$, which show no differences in this specific case due to the high effective dimension of the basal structures of the network (see red and blue curves in Figure \ref{HM-CP}(b)). Finally, Figure \ref{HM-CP}(c) shows the partition stability index for the different gaps of the dendrogram, together with the network division into four modules and the adjacency matrix. 

\subsection*{Dorogovtsev-Goltsev-Mendes graph} \label{DMG}

Dorogovtsev, Goltsev, and Mendes \cite{Dorogovstev2002} have introduced a hierarchical scale-free network in a way reminiscent of exact fractal lattices. The DGM network is a pseudofractal: it contains subnetworks resembling the whole network but lacks the affine transformation of scale. As a result, the DGM network has infinite dimensionality, containing numerous loops and being very far from tree-like. In particular, the average clustering coefficient of the network for the infinite graph is $C_{c}=\frac{4}{5}$,  a property that is suggestive of a modular organization.

\begin{figure}[hbtp]
    \centering
    \includegraphics[width=1.0\columnwidth]{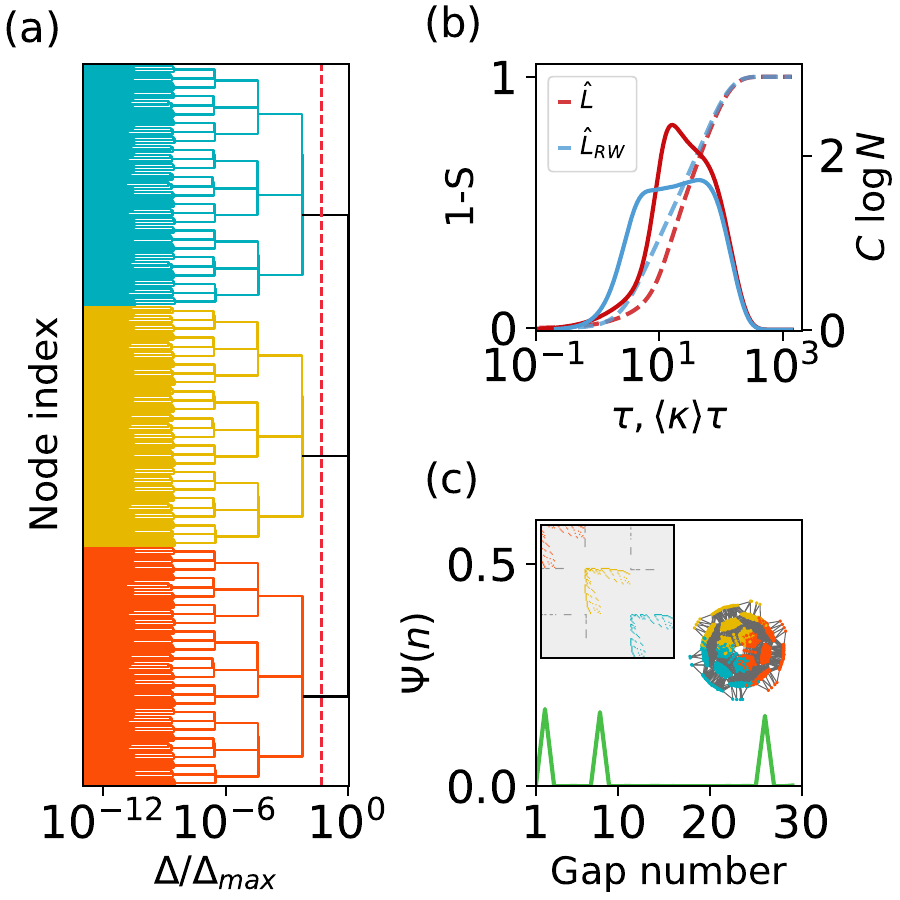}
    \caption{\textbf{Dorogovtsev-Goltsev-Mendes graph. (a)} Normalized dendrogram for a DGM network using $\tau'=1/\lambda_{\max}$. Red dashed line reflects the network division using the second gap of $\Psi$. Different communities are shaded in different colors. \textbf{(b)} Entropy parameter (dashed lines, $(1-S)$), and specific heat (solid lines, $C$), versus the temporal resolution parameter of the network, $\tau$. \textbf{(c)} Partition Stability Index $(\Psi)$ versus gap number for $\tau=\tau^\prime$. Note how peaks of $\Psi$ are equally high, reflecting the precise hierarchical structure of the network. Insets: Adjacency matrix and division into three communities of the network.}
    \label{DM}
\end{figure}

\begin{figure}[hbtp]
    \centering
    \includegraphics[width=1.0\columnwidth]{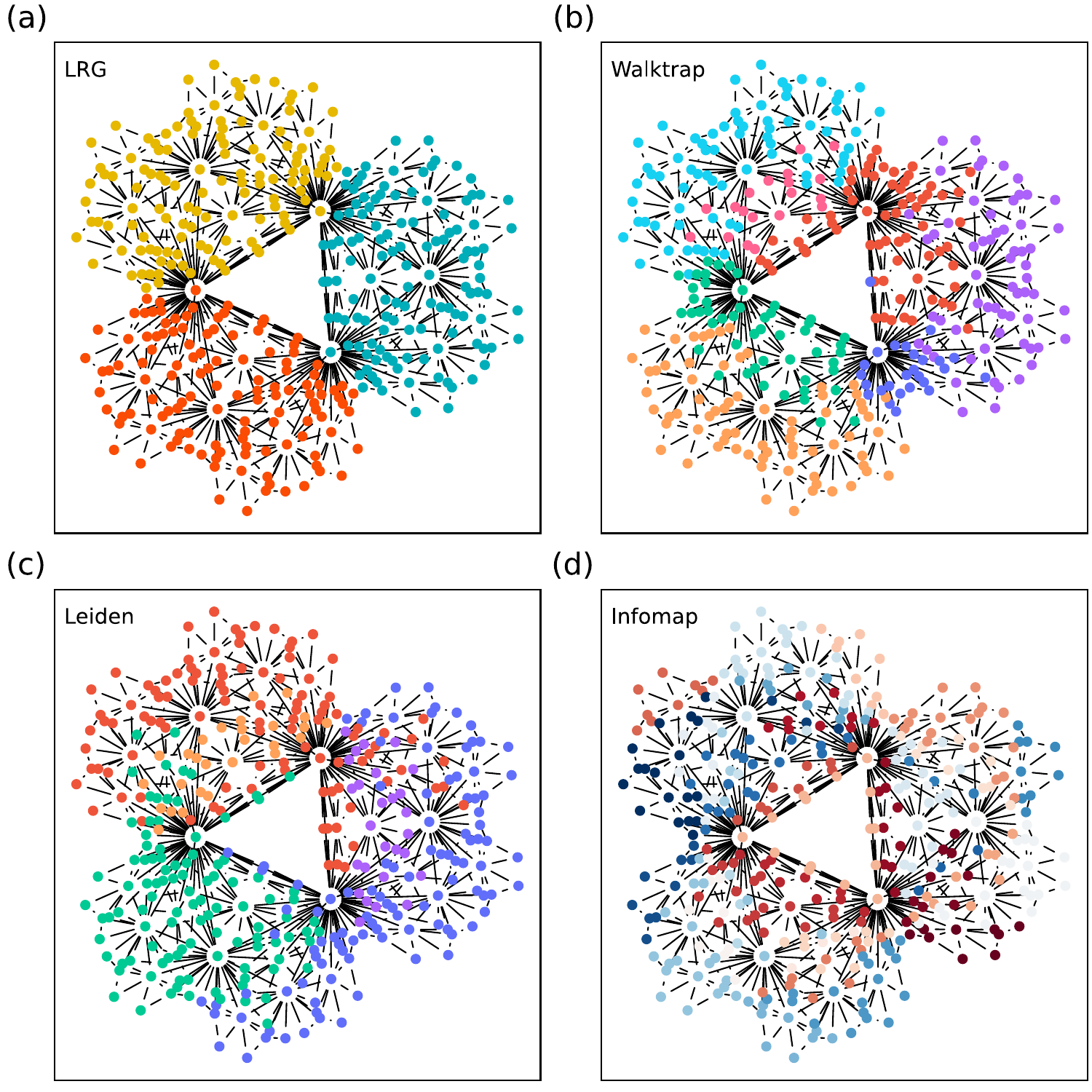}
    \caption{\textbf{Community detection methods.} Comparison of different community detection methods when applied to the DGM network: (a) MSL as stated in the previous example, (b) Walktrap algorithm (with $n_{steps}=10^3$), (c) Leiden algorithm (using $\gamma=0.004$), and (d) Infomap (using $10^2$ trials to perform the network partition). Note the accuracy of the MSL to perform network partitions in this specific case.}
    \label{DM2}
\end{figure}

Figure \ref{DM}(a) illustrates the dendrogram for a specific network and the regular nested nature of the different modules, which present the expected aggregation in the different hierarchical scales. Figure \ref{DM}(c) shows the partition stability index for the different gaps of the dendrogram, together with the network division into three modules and the adjacency matrix. Let us remark the serious issues presented by the different usual community detection algorithms when dealing with these particular type of hierarchical organization of modularity (see Fig. \ref{DM2}). We test a set of different algorithms that have proven to show an excellent performance: the walktrap algorithm \cite{Walktrap}, the Leiden algorithm \cite{Leiden} and Infomap \cite{InfoMap}. Despite this, they exhibit a great variability depending on parameters: they do not \pv{accurately predict} the different modules in the DGM network. \pv{Hence, even if it is still possible, doing some fine-tuning, to find the same communities as those of specific cuts in Figure \ref{DM}(a), we highlight that a reliable detection of the modular structure of the network is not possible without any prior knowledge of what is being searched for. Thus,} we propose the MSL as a way to go beyond the previous attempts \pv{to solve two critical issues}: \pv{On the one hand, this considers all the powers of the Laplacian needed to recover the proper network structure, giving place to a noticeable quantitative improvement on the quality of the network subdivision (see Fig. \ref{DM}(a) and Fig. \ref{DM2}(a)). On the other hand, it does not consider any topological null model that can critically constrain the possible final division into different communities, especially in this case.}

\subsection*{Lancichinetti–Fortunato–Radicchi benchmark }
\label{LFRap}

There is one further empirical test that we can make to properly apply our framework by using realistic benchmarks for community detection that account for the heterogeneity of degree and community size. In particular, the Lancichinetti–Fortunato–Radicchi benchmark considers the degree and the community size distributed as power laws. It constitutes a much harder test for algorithms and makes it easier to disclose their limits (we refer to the original work for further details on the different steps to generate the networks \cite{LFR}). Again,  this class of networks represents a challenging task, even for well-known community detection algorithms \cite{LFComm}. In particular, we choose the set of parameters $N=500$, $\tau_1=2.0$, $\tau_2=1.5$, $\mu=0.1$ and fix $\kappa_{\min} =2$ and $\kappa_{\max} =50$. First, we stress that further analysis of these benchmark networks based on multiple peaks on the entropy, together with an in-depth examination of the parameter space, can also help to understand when these networks exhibit or not communities, as a function, e.g., of the mixing parameter \cite{LFComm}. This problem will be tackled elsewhere.

\begin{figure}[hbtp]
    \centering
    \includegraphics[width=1.0\columnwidth]{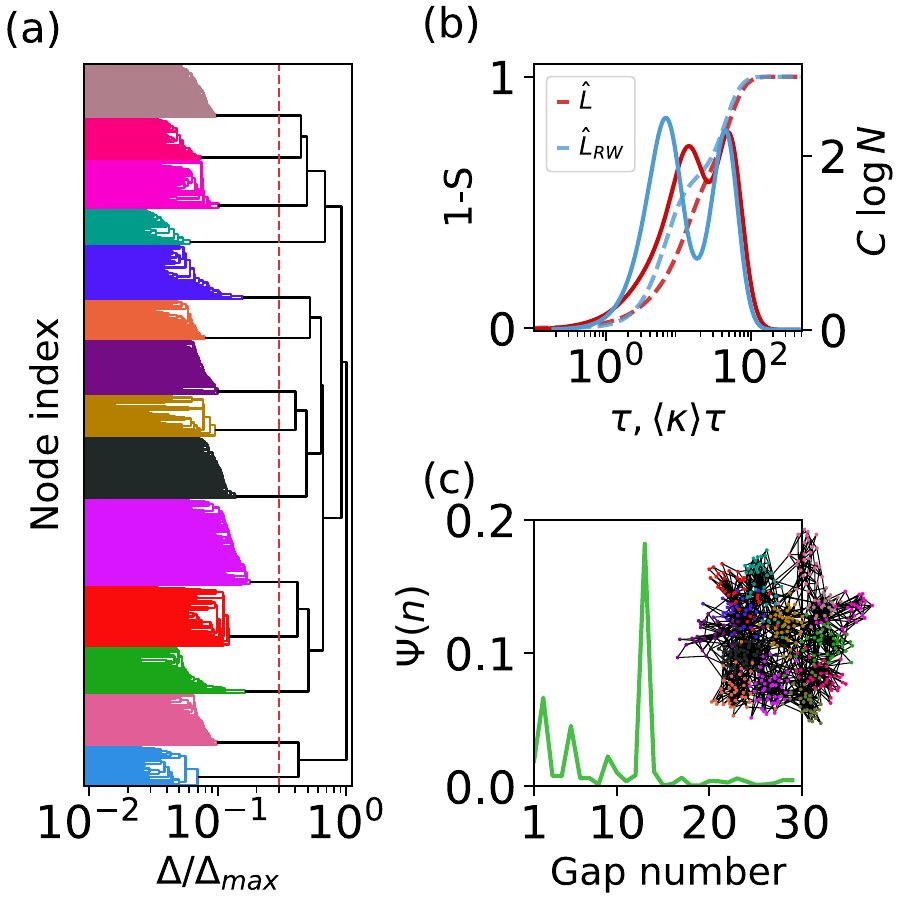}
    \caption{\textbf{Lancichinetti–Fortunato–Radicchi benchmark. (a)} Normalized dendrogram for a LFR network using $\tau=2$. Red dashed line reflects the network division using the optimal gap of $\Psi$. Different communities are shaded in different colors. \textbf{(b)} Entropy parameter (dashed lines, $(1-S)$), and specific heat (solid lines, $C$), versus the temporal resolution parameter of the network, $\tau$. \textbf{(c)} Partition Stability Index $(\Psi)$ versus gap number for $\tau=2$. Insets: Network division into communities as set by the dendrogram.}
    \label{LFR-Fig}
\end{figure}

Figure \ref{LFR-Fig} shows that the particular application of the our framework can precisely predict the predefined community of each node. Note the existence of two peaks when the spectral entropy is analyzed in this case (see Figure \ref{LFR-Fig}(b)), thus ensuring the presence of two well-defined network scales. We stress that this benchmark generates a “flat” community structure without hierarchies \cite{LFComm}, fully justifying this feature. In our view, the other significant result is the sharp double-peaked structure that emerges when we use $\hat L_{RW}$, resulting from the conceived trapping-build algorithm that generates the networks (this is yet another example of different phenomenology between $\hat L$ and $\hat L_{RW}$). Finally, we illustrate the network communities in Figure \ref{LFR-Fig}(c).

\section{The Laplacian Random-Walk}
\label{LRWAp}
It has to be reminded that $\hat L_{RW}$ is the identity matrix minus the Markovian stochastic matrix $\hat D^{-1}\hat A$ governing the time evolution of the probability distribution $\mathbf{p}_{RW}(\tau)$ to find the walker at an arbitrary vertex of the network at time $\tau$ \cite{Chung1997}. Indeed, the discrete-time random walk equation reads
\[
\mathbf{p}_{RW}(t+1)=\mathbf{p}_{RW}(t)\hat D^{-1}\hat A\,,
\]
which we can be rewritten in the continuous-time limit as
\[
\dot{\mathbf{p}}_{RW}(\tau)=-\mathbf{p}_{RW}(\tau)\hat L_{RW}
\]
Even though it is not symmetric, all eigenvalues of $\hat L_{RW}$ are real and satisfy $0\le \mu_i\le 2$ (with $=2$ only in the case of bipartite networks, which we do not consider here). It shares the spectrum with the normalized symmetric Laplacian, $\hat L_{sym}=\hat D^{-1/2}\hat L\hat D^{-1/2}$ as they are related by the similarity transformation $\hat L_{sym}=D^{1/2}\hat L_{RW}\hat D^{-1/2}$. The respective eigenvectors, sharing the same eigenvalue, are related by the simple transformation $\mathbf{v}_{RW}=\hat D^{-1/2}\mathbf{v}_{sym}$.

\begin{figure}[hbtp]
    \centering
    \includegraphics[width=0.9\columnwidth]{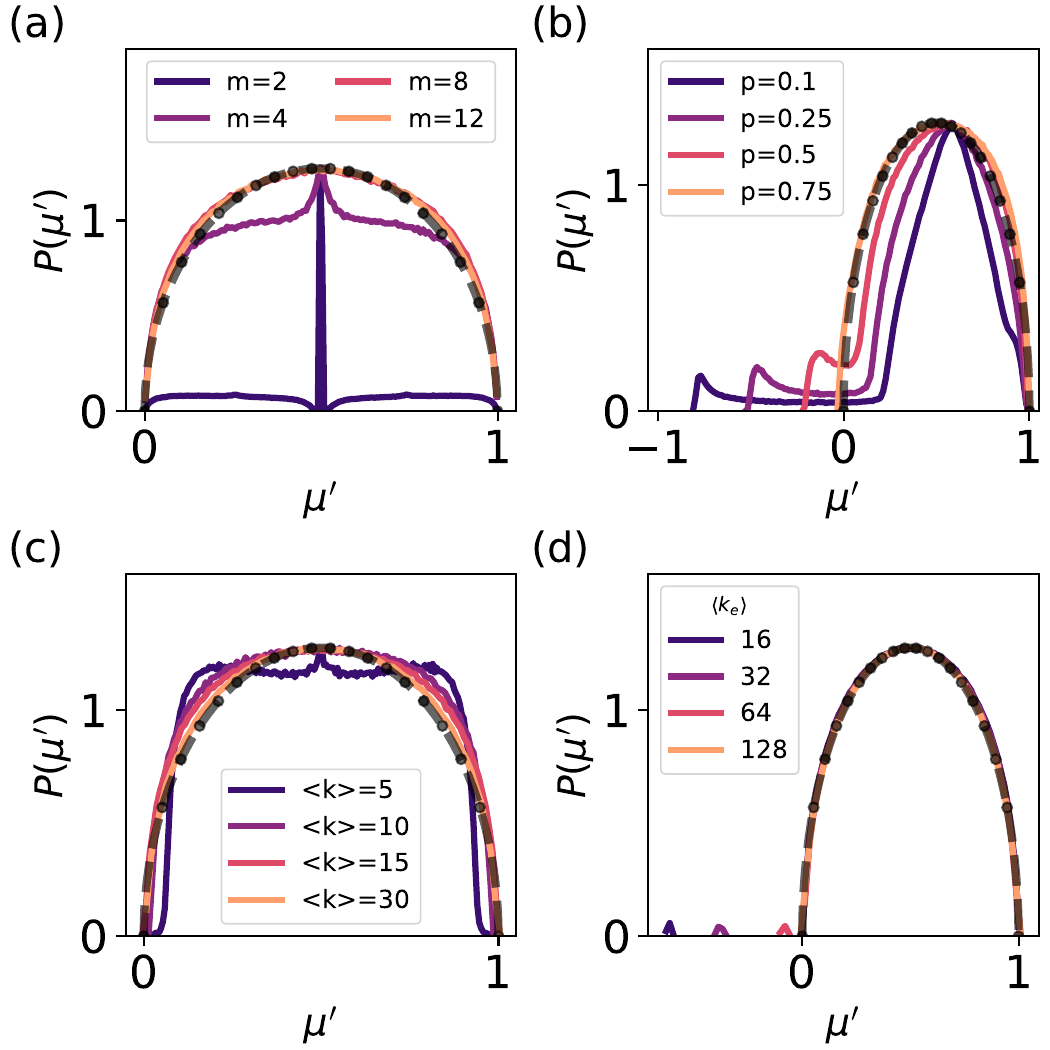}
    \caption{\textbf{Laplacian Random-Walk.} Probability distribution of eigenvalues, $P(\mu')$ for $\hat L_{RW}$ and different networks: \textbf{(a)} Barabasi-Albert, varying $m$ (see legend). \textbf{(b)} Watts-Strogatz with $\langle \kappa \rangle=18$, varying $p$ (see legend). \textbf{(c)} Giant connected component of Erd\H{o}s-R\'enyi networks varying the mean connectivity $\langle \kappa \rangle$ (see legend). \textbf{(d)} Stochastic block model with four modules of equal size, with internal connectivity $\langle \kappa_i \rangle = 128$, varying the external connectivity between blocks ($\langle \kappa_e \rangle$, see legend). Parameters: $N=2048$. All networks have been averaged over $10^2$ independent realizations.}
    \label{RWApp}
\end{figure}

In particular, $\hat L_{RW}$ also characterizes the return-time distribution of the random walker: a walker starting at $\tau=0$ from an arbitrary node of the network has a probability $P_0(\tau)$ of returning to the initial node. Therefore, by using the usual Laplace transform, $P_0(\tau)=\int d\mu\, p(\mu)e^{-\mu \tau}$, directly linking the spectral density $p(\mu)$ of $\hat L_{RW}$ with the probability density function of return time distributions in any network \cite{Barrat2008}.

We here introduce the following change of variables to split the spectrum of eigenvalues in two parts, as indicated in the main text: $\mu'=\frac{1}{2}+\frac{\mu-1}{2(\mu_{\max}-1)}$, with $-1\le \mu_i\le 1$. Note that as  the system is ergodic $\mu=1$ is always a non-degenerate eigenvalue and that $\mu_{max}\le 2$ with $\mu_{max}=2$ only for bipartite networks. In particular, $\mu'_{max}=1$ represents now an upper limit for all networks and the spectral density can by redefined as $P(\mu')=2(\mu_{\max}-1)p[2-\mu_{max}+2(\mu_{\max}-1)\mu']$. 

Figure \ref{RWApp} shows the convergence, for diverse selected networks, of the spectral density distribution, $P(\mu')$ to the theoretical --and universal-- Wigner semicircle law \cite{Chung2003, Dorogovtsev2022}. 
%Note that a similar result has been obtained for the mean-field theoretical expectation before $x(t)$ first returns at time $T$ to its initial value $x(0)$; i.e., $\langle x(s) \rangle=\frac{8}{\pi}\sqrt{s(1-s)}$, with $s=t/T$ \cite{Baldassarri2003}. 
We also highlight that negative values of $\mu'$ reflect multi-scale effects of the system.

\section{Effects of $\hat L$ and $\hat L_{RW}$ in the Star-Clique network}
\label{CST}

\begin{figure}[hbtp]
    \centering
    \includegraphics[width=0.9\columnwidth]{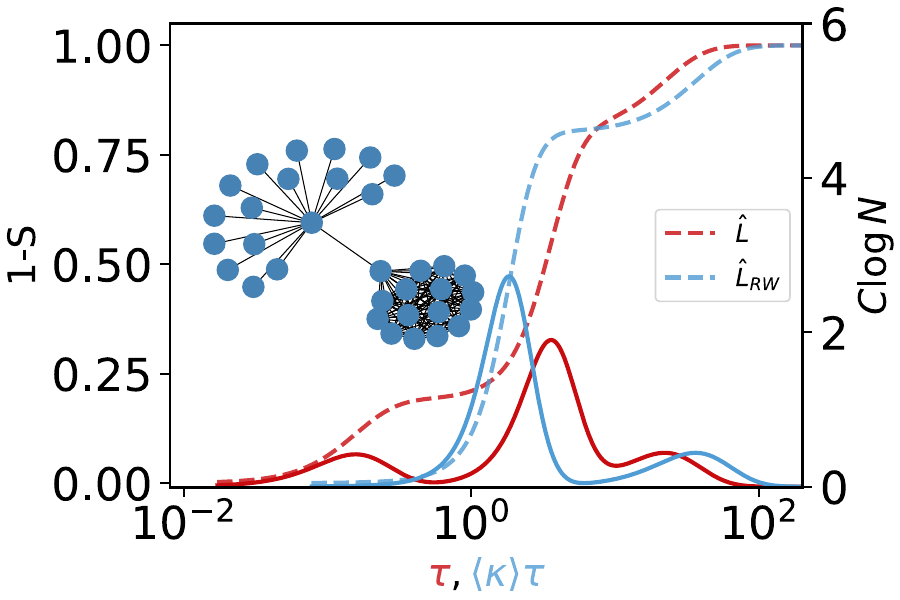}
    \caption{\textbf{Star-Clique network.} Entropy parameter (dashed lines, $(1-S)$), and specific heat (solid lines, $C$), versus the temporal resolution parameter of the network, $\tau$, for both Laplacians, $\hat L$ and $\hat L_{RW}$. Note the explicit differences in the total number of peaks of the specific heat because $\hat L_{RW}$ reflects the trapping time into the two main communities of the network: the star on the one hand and the clique on the other.}
    \label{CSTAR}
\end{figure}

As previously shown in the main text, both $\hat L$ and $\hat L_{RW}$ can be used to detect emergent communities in a network. The main difference is that while $\hat L$ describes a continuous flow typical of the heat equation uniformly flowing \pv{through} each link, $\hat L_{RW}$ encodes the Markovian evolution of the probability distribution on the network vertices of a random walker throw a transition matrix that imposes a normalized and uniform distribution of single step transitions from each starting vertex to neighbors. This makes the flow between two weakly connected subnetworks with a high density of internal connections very slow. This effect is well captured  by \pv{applying} of our community detection algorithm with both Laplacians in the Star-Clique network illustrated in Figure \ref{CSTAR}. 

The Star-Clique network is formed by a dense module of interconnected nodes connected to a star graph within one of the clique nodes. This serves as a powerful example of the main differences between both Laplacian matrices. On the one hand, the Laplacian, $\hat L$, reflects three main scales in the three peaks of the specific heat: one emerging when the clique is integrated out, one emerging when the star is integrated out and, finally, one corresponding to the full network structure (taking into account two modules and two hubs) completely integrated out, looking at the network as a single node. On the other hand, $\hat L_{RW}$, reflects two main scales due to the trapping nature of the RW process encoded in the matrix: the first peak \pv{captures} the existence of two essential communities, which are finally integrated out in the last peak of $C$.

%\section{Metastable nodes}

%\subsection*{Zachary's Karate club}

\def\url#1{}
%\bibliography{Modularity}

\end{document}